\documentclass[12pt,preprint]{aastex}
\newcommand\cf{cf.}
\newcommand\ie{i.e.}
\newcommand\eg{e.g.}

\newcommand\etal{et al.}
\newcommand\vs{vs.}

\newcommand\chandra{\textit{Chandra}}
\newcommand\rosat{\textit{ROSAT}}
\newcommand\xmm{\textit{XMM-Newton}}
\newcommand\msun{M_\odot}

\newcommand\nelec{n_{\rm e}}
\newcommand\nh{n_{\rm H}}
\newcommand\heat{{\cal H}}
\newcommand\cool{{\cal R}}
\newcommand\tcool{t_{\rm c}}
\newcommand\theat{t_{\rm h}}
\newcommand\bmug{B_{\mu\rm G}}
\newcommand\mdotbondi{\dot M_{\rm B}}
\newcommand\entind{K}  % For Mark

\shorttitle{Mechanical AGN Feedback}
\shortauthors{McNamara \& Nulsen}

\begin{document}


\title{Mechanical Feedback from Active Galactic Nuclei in Galaxies, Groups,
  and Clusters} 


\author{B. R. McNamara\altaffilmark{1,2,3,4} \&
  P. E. J. Nulsen\altaffilmark{3,5}}  


\altaffiltext{1}{Department of Physics \& Astronomy, University of
  Waterloo, Waterloo, ON, Canada} 
\altaffiltext{2}{Perimeter Institute for Theoretical Physics,
  Waterloo, ON, Canada}  
\altaffiltext{3}{Harvard-Smithsonian Center for Astrophysics, 60
  Garden St, Cambridge, MA 02138}
\altaffiltext{4}{email: mcnamara@uwaterloo.ca}
\altaffiltext{5}{email: pnulsen@cfa.harvard.edu}

\begin{abstract}
The radiative cooling timescales at the centers of hot atmospheres surrounding
elliptical galaxies, groups, and clusters are much shorter than their
ages.  Therefore, hot atmospheres are expected to cool and to form
stars.  Cold gas and star formation are observed in central cluster
galaxies but at levels below those expected from an unimpeded cooling
flow.  X-ray observations have shown that wholesale cooling is being offset by
mechanical heating from radio active galactic nuclei.
Feedback is widely considered to be an important and perhaps unavoidable
consequence of the evolution of galaxies and supermassive black holes.
We show that cooling X-ray atmospheres and the ensuing star formation
and nuclear activity are probably coupled to a self-regulated feedback loop.
While the energetics are now reasonably well understood, other aspects of
feedback are not.  We highlight the problems of
atmospheric heating and transport processes, accretion, and nuclear
activity, and we discuss the potential role of black hole spin.  We
discuss X-ray imagery showing that the chemical elements produced by
central galaxies are being dispersed on large scales by outflows launched from
the vicinity of supermassive black holes.  Finally, we comment on the growing evidence
for mechanical heating of distant cluster atmospheres by radio jets
and its potential consequences for the excess entropy in hot halos and
a possible decline in the number of distant cooling flows.

\end{abstract}

\keywords{galaxies clusters: general --- intergalactic medium ---
  X-rays: galaxies: clusters}

\section{Introduction}

The large scale structure of the Universe revealed by WMAP
\citep{sbd07, ksd11}, distant supernovae \citep{pag99, rfl00},
% Not the first paper by Perlmutter, but the one with the main
% result.  Also, Schmidt (1998) says it can be done, but the result is
% in Riess.
and galaxy clusters \citep{vkb09, mar10,aem11} agrees remarkably well with
the $\Lambda$CDM cosmogony \citep{def85}.
However, additional physics is needed to explain the numbers and distribution of 
baryons in galaxies and galaxy clusters.  Gas dynamical models of dark
matter halos incorporating radiative cooling and gravitational heating
alone produce too much cold gas, too many young stars, and too few hot
baryons \citep{bpb01, dro01, b07}. % One of the most important issues concerns
% radiative cooling, a process that becomes
% increasingly efficient with increasing gas density.  
The level of overcooling leads to model galaxy
luminosities, colors, disk and bulge sizes, and so forth 
that fail to match observations.  A similar problem is found on
larger scales.   Hydrodynamic models have difficulty
reproducing the central gas densities, temperatures, entropy
distributions, and baryon fractions of the hot atmospheres of clusters
\citep[for excellent reviews, see][]{v05, bk09}.

\begin{figure}[!ht]
\begin{center}
\includegraphics[height=0.35\textheight]{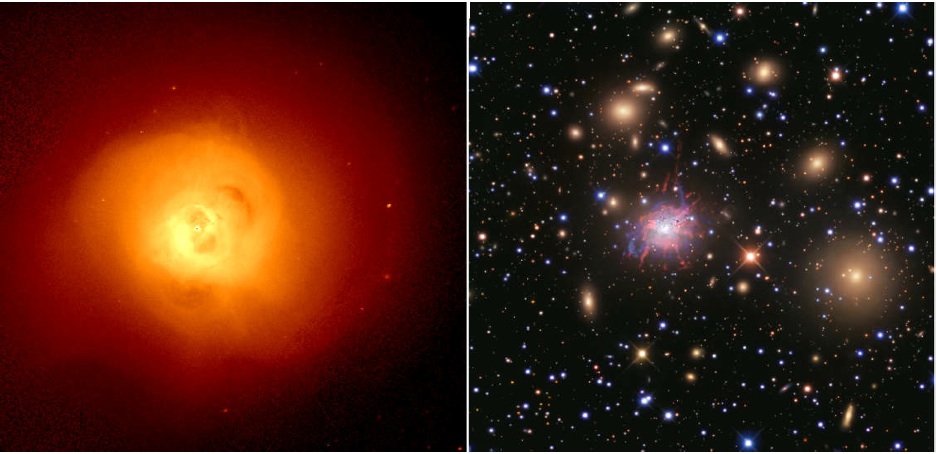}
\end{center}
\caption{\textit{Left:} Deep \chandra{}  X-ray image of the Perseus
  cluster \citep{fsa11}.  \textit{Right:} Matched optical image of Perseus
  cluster showing the extensive system of line-emitting filaments
  around NGC~1275 \citep{fsa11}. The images are 11.8 arcmin from N to
  S. } \label{fig:perseus} 
\end{figure}

The formation and evolution of the stars and gas (baryons) are more
complicated than the evolution of dark matter halos.  Dark
matter is governed almost exclusively by gravity.  Baryons respond to
more complex processes, such as heating, cooling, transport processes,
and energetic feedback from supernova explosions and active galactic nuclei (AGN).  
Evidently, outflows and winds powered by supernovae and
by accretion onto supermassive black holes (SMBHs) play a significant
role in the evolution of galaxies at essentially all stages of their
development \citep[for reviews see][]{ah11, vcb05}.   For example, energetic outflows have been detected using emission and absorption line features from distant radio
galaxies \citep[e.g.,][]{nle06, mhs07, ltn11}, quasars \citep[e.g., ][]{dkab01, cbg07, rv11}, and starburst galaxies \citep{ham90, hls00, martin05, ass10}.  Quasar winds
capable of sweeping galaxies of their gas may be the
primary agent driving the coevolution of massive galaxies and SMBHs
\citep{sr98, dsh05}. Star formation in galaxies lying along the
so-called main sequence of galaxy formation \citep{nwf07,tgn10} at redshift
$z\sim2$ and below
% (i.e., cosmic downsizing \citep{csh96}) 
may be maintained by outflowing gas launched
by supernova explosions and inflowing cold accretion flows
\citep{fd08, dob11}.  Supernova and AGN
feedback may drive the rapid evolution of
galaxies from the blue to the red sequences \citep{bgb04,sts07, hjf09}. 
Finally, AGN feedback is probably quenching cooling flows in galaxies and clusters
and preventing star formation.  

The theory of galaxy formation has advanced significantly with the addition of simple feedback prescriptions into traditional semianalytic and hydrodynamic 
galaxy formation models \citep[e.g.,][]{kwg93, sp99, baugh06, sh03}.  Recent studies have shown that a plausible combination of supernova and AGN feedback is able regulate the growth of galaxies by heating or expelling gas from halos \citep{ss06, bbm06,csw06, bmb08}.  
A triumph was achieved when models incorporating radio (mechanical) AGN feedback occurring at late times reproduced the observed luminosity function of massive galaxies \citep{bbm06, csw06, dfo11,gdo11}.   Modeling the luminosity function of galaxies while simultaneously reproducing the observed properties of the hot atmospheres surrounding them is a further challenge \citep{bmb08,msp10}.  
%While the connection between radio AGN feedback and the maintenance of red and dead ellipticals is now well established, a relationship between feedback
%and the early evolution of galaxies is largely circumstantial.

Evidence for a genuine AGN feedback cycle has emerged
from X-ray observations of the hot atmospheres of galaxy clusters.
A wealth of data collected by the \xmm{}
and \chandra{} X-ray observatories has shown that radio AGN are probably
the principal agent  heating the hot
atmospheres of galaxies, clusters, and groups and suppressing cooling flows.
%A sketch of this process follows: 
The process works roughly as follows: The atmosphere cools and condenses into molecular clouds that form
stars and feed the supermassive black hole concealed within the BCG.
Nuclear accretion and black hole spin produce
mechanically-powerful radio AGN that heat the cooling atmosphere, slowing
the rate of cooling, and the cycle repeats. 
%, which is conceptually simple but physically complex,
%probably describes primeval galaxy formation.
Star formation is suppressed almost entirely in elliptical galaxies, keeping them ``red and dead."
%A triumph was achieved when models incorporating radio (mechanical) AGN feedback occurring at late times reproduced the observed luminosity function of massive galaxies \citep{bbm06, csw06, dfo11,gdo11}.   Modeling the luminosity function while simultaneously reproducing the observed properties of the hot atmospheres surrounding them is a further challenge \citep{bmb08,msp10}.  

\begin{figure}[!ht]
\begin{center}
\raisebox{0.025\textheight}{%
\includegraphics[height=0.2\textheight]{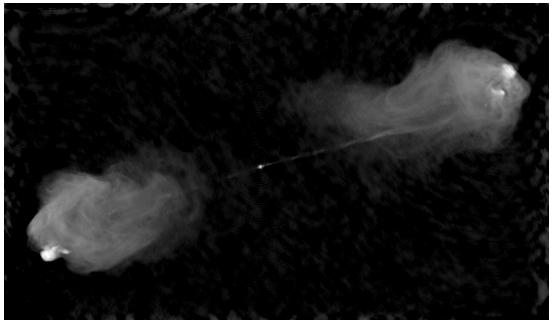}}
\includegraphics[height=0.25\textheight]{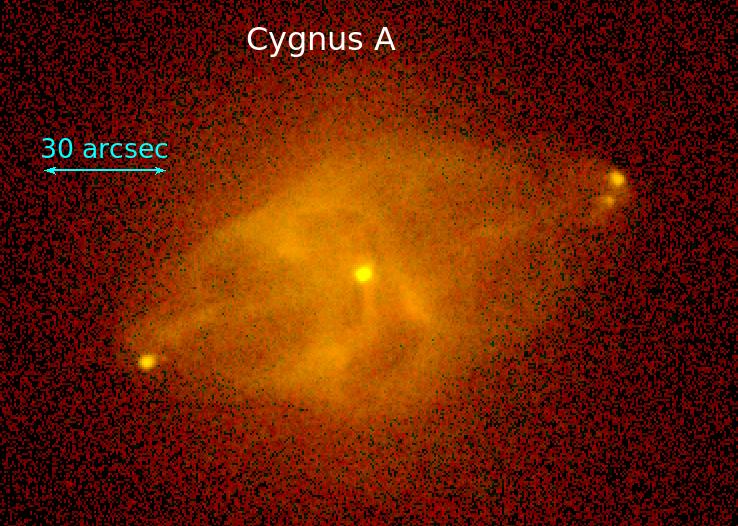}
\caption{\textit{Left:} Cygnus A, 6 cm VLA radio image \citep{pdc84}.
  \textit{Right:} \chandra{} X-ray image of Cygnus A.  The AGN and
  radio hotspots are visible in both images.  Complex structure,
  including cavities and the cocoon shock can be seen in the X-ray
  image \citep{swa02}.  }
\label{fig:cyga}
\end{center}
\end{figure}

Here we discuss the primary observational evidence for this cycle,
concentrating on the spectacular X-ray images of cavities, shock
fronts, and filaments embedded in hot atmospheres
(Figs.~\ref{fig:perseus} and \ref{fig:cyga}). We briefly discuss
observations at longer wavelengths that track the cooling gas and star
formation through the final phase of the feedback cycle.
We comment on some of the theoretical problems challenging a more complete
understanding, including how the central engine is powered, how jet
power is coupled to the surrounding hot atmospheres, and recent
advances in plasma physics.  We discuss new evidence for
metal-enriched outflows, and the growing literature dealing with AGN
feedback in distant clusters.  

This Focus Issue brings together articles dealing with the physics of
galaxy clusters, several of which are closely related to ours.  In
keeping with the spirit of this volume, we make no attempt to present
a comprehensive review of what in recent years has grown to a vast
literature.  Instead we focus on several topics of interest to us and
refer the reader to other contributions to this volume and to several
comprehensive reviews on related topics: cooling flows \citep{f94},
elliptical galaxies \citep{mb03}, 
X-ray clusters and cosmology \citep{v05,aem11}, X-ray spectroscopy of
clusters \citep{ pf06}, feedback in X-ray clusters, groups, and
galaxies \citep{mn07, bcs10, gbm11}, extragalactic radio jets
\citep{w09}, feedback in galaxies \citep{cfb09}, and the outstanding and
comprehensive review of the history of black holes and AGN in the
Universe by \citet{ah11}.
%Further discussions of these and closely related topics can be found in articles appearing in this
%volume by Ruszkowski, Fabian, Jones, Sun, and McCarthy [editor, please
%  check].

\section{X-ray Cavities as Gauges of Jet Power} \label{sec:gauges}

The radiative cooling time at the centers of hot atmospheres in
groups and clusters is often less than 1 Gyr.  In elliptical
galaxies it lies below 0.1 Gyr.  Unless the thermal energy
being radiated away is replenished, the gas will cool and accrete onto
the central galaxy and form stars \citep{f94}.  The end products of cooling
in the forms of cold molecular clouds
and star formation are observed in many BCGs \citep{jfn87, e01, sc03,
  obp08}, but at levels far below those expected from persistent cooling
over the ages of clusters.  Other potential repositories include nearly
invisible low-mass stars \citep{fnc82, so83, j86} or neutral hydrogen clouds
spread throughout clusters \citep{hbr78}.  The failure to find a
long-term repository in
cold gas and stars led many to conclude that very little gas is
actually cooling to low temperatures.  Several mechanisms that would
heat the gas and prevent it from cooling have been suggested over the years, including
dynamical friction \citep{miller86}, thermal conduction from the hot
outer atmospheres of clusters \citep{rt89,bd88}, and  AGN \citep{pgd90, td97, swd01, bt95}.  However, atmospheric heating models have only recently gained traction.  

 X-ray images of the Perseus and Cygnus A clusters taken with the \rosat{} observatory  were the first to show radio AGN interacting strongly with the hot atmospheres surrounding them \citep{bvf93, cph94}.  Cavities and filamentary structure in the atmospheres of both clusters
 are located near the 
 central radio sources.  Discoveries of structure in the atmospheres
of other clusters soon followed \citep{sbr95, rlb00}.   With its leap in spatial and spectral resolution, the \chandra{} X-ray observatory revealed cavities, shock fronts, and cool filaments near the central radio sources in essentially all cooling flow clusters it has observed. Noting the close association between the radio lobes and X-ray cavities
in Hydra A, \citet{mwn00} pointed out that cavity volumes multiplied by their
surrounding pressures provide a gauge of the $pV$ work (mechanical energy) expended as
the cavities are inflated.
Assuming the cavity dynamics  are governed by buoyancy, their ages and in turn mean
jet power, were estimated \citep{mwn00, csf02, brm04}. \chandra{} has since
provided what are arguably the most reliable measurements of jet power in dozens of galaxies
and clusters.  These measurements have led to the understanding that cooling atmospheres
in clusters and giant elliptical galaxies are regulated by AGN feedback.

\begin{figure}[!ht]
\begin{center}
\includegraphics[height=0.35\textheight]{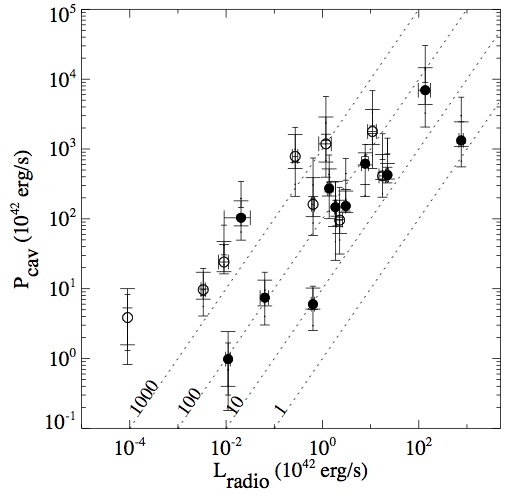}
\includegraphics[height=0.35\textheight]{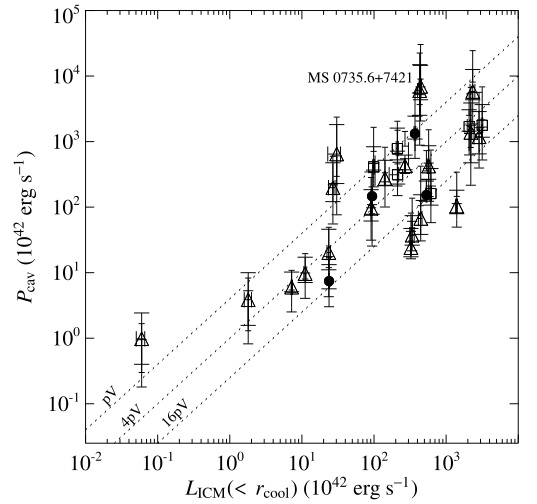}
\caption{\textit{Left:}  Cavity power \vs{} broadband radio power for
  a sample of cool core clusters \citep{bmn08}.  \textit{Right:}
  Cavity power \vs{} X-ray power radiated from the region where the
  cooling time is shorter than 7.7 Gyr\citep{rmn06}.
}
\label{fig:heatcool}
\end{center}
\end{figure}

The total enthalpy, \ie, the $pV$ work plus the internal energy that
provides the pressure supporting the cavities, is given by
\begin{equation}
H = {\gamma \over \gamma - 1} pV = \cases{2.5 pV,&for $\gamma = 5/3$\cr 
  4pV,&for $\gamma = 4/3$\cr},
\end{equation}
where $\gamma$ is the ratio of specific heats of the cavity plasma.
The appropriate ratio of specific heats depends on whether the
pressure support within cavities is supplied primarily by relativistic
plasma ($\gamma = 4/3$) or nonrelativistic plasma ($\gamma = 5/3$).
Were the radio lobes filled with a dilute thermal
plasma in local pressure equilibrium, its temperature must
exceed roughly 20 keV in order to avoid detection by its thermal X-ray
emission \citep{bsm03,gmn07}.  Such a plasma would presumably have
been heated by shocks or cosmic rays stemming from the AGN outburst.
% However, shock strengths in
% clusters are probably too weak to heat the gas to such high
% temperatures.  
More likely, the cavities are filled with relativistic plasma
from the radio jets yielding $H = 4pV$ per cavity.
Of course, cavities may be filled with more complex gas mixtures.
Ideally, the prefactor multiplying $pV$ would take full account of the
work done by an expanding cavity (the energy radiated is generally
negligible).  Using the enthalpy allows  $1pV$ for the work.
But since the pressure confining an expanding cavity probably
decreases as it expands, this is generally an underestimate.  The work
done depends on how quickly the cavities were inflated and, more
generally, the history of each AGN outburst.  Observations select for
large, mature, cavities such that the canonical $4pV$ per cavity is
roughly consistent with MHD simulations \citep{moj11}.  The $4pV$ approximation
may be less accurate for cavities inflated by cosmic rays \citep{mb08}. 
ALMA observations of the Sunyaev-Zeldovich effect, which is most sensitive to nonrelativistic
electrons \citep{pes05}  will help to determine the composition of the
lobe plasma.
% Relativistic electrons scatter and shift photon energies, but,
% because they produce large energy shifts, the effect stops
% increasing with the mean electron energy (declines roughly as
% 1/gamma, if the electrons still make a significant contribution to
% the pressure).

The synchrotron power of radio jets and lobes represents a minor
fraction of their total power \citep[\eg,][]{s74}.  It should be no
surprise that radio AGN, which are observed in most BCGs centered in
cooling flows,
% \citep{b90, bvk07, cdv08} 
were long ago suspected to
play a significant role in quenching cooling flows
\citep[\eg,][]{b90,pgd90, bo91, tb93, td97}.  However,
radio-based estimates of their mechanical power relied critically on the unknown
ratio, $k$, of the energy in non-radiating particles to that in the
synchrotron emitting electrons, assumed to lie in the range $0-100$
\citep[see][]{wrb99}.
%% Nonradiating particles could easily be thermal (including
%% electrons). 
A low value of $k$ would provide too little mechanical energy to
offset cooling; higher values were possible but received little
observational support.

As a consequence of  high
atmospheric pressures, $\sim 10^{-10}\rm\ dyne\ cm^{-2}$, and large cavity diameters, typically $20$ kpc but upward of $200$ kpc in cases, \chandra{} revealed that relatively modest radio sources may carry extraordinarily large
mechanical powers.  Fig.~\ref{fig:heatcool} \textit{left} shows a plot
of cavity power against bolometric radio power for a sample of radio AGN in
clusters and groups \citep{bmn08}.  The diagonal lines represent fixed
ratios between cavity power and synchrotron luminosity.  On average, the mean
mechanical power is $100$ times larger than the synchrotron power, and in some 
instances $\sim 1000$ times larger.  With mechanical jet power exceeding $10^{45}\rm ~
erg~s^{-1}$,  many radio sources rival the radiative output of a quasar.  Their mechanical
power is comparable to or exceeds the X-ray luminosities of their cooling atmospheres.
% At the same time,
%their synchrotron luminosities of $\sim 10^{42}~\rm erg~s^{-1}$ are
%modest compared to the X-ray luminosities of their cooling cores.
%Relatively weak radio sources in cluster cores can be mechanically
%powerful enough to quench cooling flows.  
The high mechanical (jet) powers relative to their synchrotron luminosities implies that
hydrodynamical radio sources are governed on large scales by heavy
particles, \ie, $k \gg 10$ \citep{df06, bmn08, deyoung06, hc10}. 
If jets are launched as light particles or Poynting flux  \citep{dlf08, deyoung06, ntl08}, their
radio lobes, which fill the cavities, must be energetically dominated by heavy particles
that were presumably entrained from the surrounding atmosphere \citep{croston08}.

%****HIGH k IN THE LOBE IS NOT THE SAME AS
%HIGH k IN THE JET.  I DO NOT FOLLOW THE INFERENCE HERE.  IT SEEMS TO
%ASSUME THAT JETS SIMPLY INFLATE LOBES, WITH NO FURTHER PROCESSING.
%WHAT IF JETS ARE POYNTING DOMINATED? --- ALSO SEE NEAR THE END OF
%SECTION \ref{sec:spinobs}****

\section{Quenching Cooling Flows} \label{sec:quench}

A significant consequence of high jet powers is their ability
to offset radiative cooling of the atmospheres
surrounding them \citep{brm04, rmn06, df06, bkh06, drm10}.  Fig.~\ref{fig:heatcool}
\textit{right} shows average cavity power plotted against the X-ray
cooling luminosity for a sample of clusters.  Lines of
equality between cooling and heating are shown for injected
energies of $1pV$, $4pV$, and $16pV$ per cavity.  The diagram shows that
$\simeq
4pV$ per cavity is typically observed, which is enough to offset cooling in most systems \citep{mn07}.
AGN power output is variable.  Objects move up or down 
in $P_{\rm cav}$ depending on when
they are observed.  Therefore, the current power,  low or high with respect
to the $4pV$ line, may not equal its long term average.

Selection effects are at issue in Fig.~\ref{fig:heatcool}.
At high cooling luminosities, the cavity detection fraction is about 70\%  \citep{df06}.  With
the general bias against detection \citep{mn07}, their result implies that nearly all strong cooling
flows harbor powerful cavity systems.  At the same time,  we are aware of no powerful systems in low cooling luminosity clusters lying to the left of the $pV$
line. They would be easily detected.  We interpret the distribution of points as an envelope sampling the upper
end of the distribution of jet powers at a given cooling luminosity, and not a linear correlation.   Most importantly, the diagram shows that jet power correlates with the level
of cooling:  larger cooling flows host increasingly powerful AGN that rival or exceed their
cooling luminosities.  

Fig.~\ref{fig:heatcool} does not include additional power from shocks \citep{dnm01}, sound waves \citep{fsa03, fnh05}, thermal conduction
\citep{zn03, vf04, v11}, cosmic ray leakage \citep{mb08}, and other forms of energy.  On average, cavity
enthalpies underestimate the energy deposited in the cooling
regions, which only strengthens our view that AGN are energetically able to offset cooling.

Fig.~\ref{fig:heatcool} is weighted toward cavity systems in rich clusters.
Cavity systems have recently been identified in abundance
in the hot atmospheres of groups \citep{svd09, drm10, gog10, gov11, ogd11}
and ellipticals \citep{cmn10,njf09, mj10}.  These studies have shown
that radio AGN in groups and ellipticals are energetically significant with respect to the thermal energy content
of their atmospheres \citep{ gbd11,msp10}.   

The upshot is AGN feedback is operating at
the centers of cooling atmospheres spanning seven decades in X-ray
luminosity between $\sim 10^{38}~\rm erg~s^{-1}$ and $10^{45}~\rm
erg~s^{-1}$.  This is significant because other heating processes such as 
thermal conduction and dynamical friction are unimportant in the atmospheres
of groups and ellipticals.   That AGN feedback is operating over such
a broad scale suggests it is the primary mechanism stabilizing cooling flows
in galaxies,  groups, and rich clusters.

\subsection{Scatter in Cavity Power vs Cooling Luminosity Diagram}

The scatter  in Fig.~\ref{fig:heatcool} \textit{right} is
largely real.  We find no correlation between this scatter and physical parameters,
including central gas and stellar densities, host galaxy luminosity,
synchrotron age, and so forth.  As deeper \chandra{} images yielding better cavity 
measurements have become available, it is
becoming clear to us that measurement
uncertainty contributes to the scatter beyond the errors shown in Fig.~\ref{fig:heatcool}.  The error
bars reflect the uncertainties in the spheroidal geometry.  In many
cases, cavity sizes and locations are measured using underexposed
\chandra{} images that are unable to reveal the extent of the cavity.
In addition, irregular shapes add uncertainty to
estimates of their volume and location within a cluster. This in turn adds
uncertainty to the estimate of the surrounding pressure.  Finally, many cavities are
surrounded by thick rims.  It is unclear whether the inner, outer, or midpoint of the
rim represents the true radius and hence, the displaced volume of the cavity.
While this technique is straightforward, unavoidable complications
contribute significantly to the uncertainty of the measurements.
 Deeper images combined with more objective
measuring techniques are needed to reduce the measurement error.

\begin{figure}[!ht]
\begin{center}
\includegraphics[height=0.35\textheight]{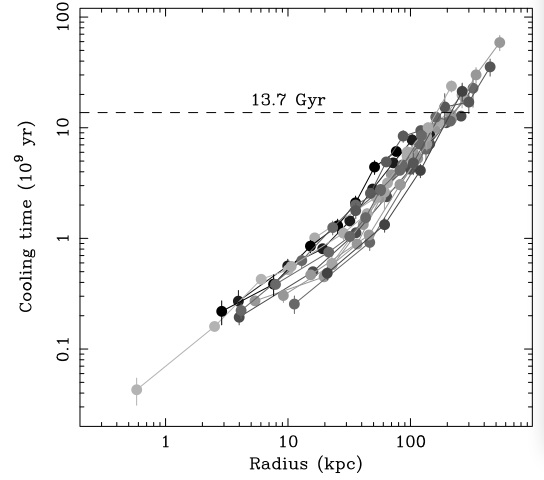}
\includegraphics[height=0.35\textheight]{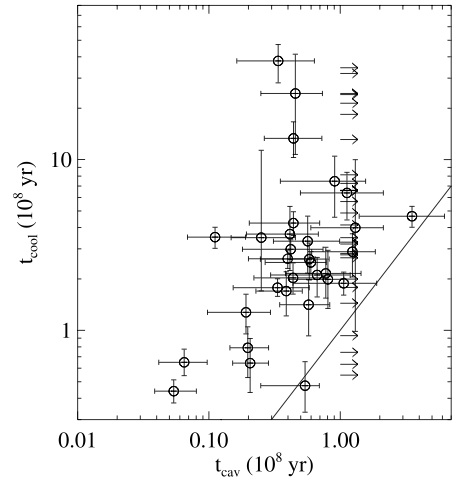}
\caption{\textit{Left:} Cooling time \vs{} radius for the sample of
  \citet{vf04}.  \textit{Right:} Cooling time \vs{} cavity age from
 \citep{rmn08}.} 
\label{fig:tcool}
\end{center}
\end{figure}

\section{Maintaining X-ray Atmospheres through Mechanical Feedback}

The list of processes unrelated to AGN that can heat the hot
intracluster gas continues to expand and it is too long to cover in
any detail here.  The most important distinction among potential
heating processes is whether or not they involve feedback.  The great
majority of proposed heating processes, including thermal conduction
\citep[\eg,][]{nm01}, turbulent heat diffusion \citep[\eg,][]{ro11},
heating associated with mergers \citep[\eg,][]{zmj10, bd11} dynamical
friction \citep[\eg,][]{m86, kek05} and heating by off center AGN
\citep{hsh09}, involve no evident feedback mechanism, hence no means
for heating rates to be tuned to match radiative cooling.  For these
to account for the observed state of clusters we must appeal to
accidental coincidence between average heating and cooling rates.

\subsection{Heating \vs{} cooling}

The terms ``heating'' and ``cooling'' still cause confusion when
applied to cool core clusters.  Radiative losses from gas in a fixed
volume would reduce its temperature.  However, in an atmosphere
confined by gravity, the heat loss due to radiation is counteracted by
adiabatic compression under the weight of overlying gas, so that the
sign of the net temperature change is indeterminate.  Under realistic
conditions, heat loss can result in a temperature rise.  Conversely,
heat input to a gravitationally confined atmosphere does not
necessarily increase the gas temperature.  Apart from any heat
exchange, other forms of energy input (\eg, the increase in potential
energy when gas is pushed outward by inflating lobes) complicate this
issue further.  Thus, ``heating'' is frequently associated with a
decrease in gas temperature, while ``cooling'' can cause the
temperature to increase.

While the indeterminate temperature change creates ambiguity, the
sense of heat transfer does not.  Here, ``cooling'' always refers to
heat loss (by radiation) from the gas.  Its invariable physical
consequence is a reduction in the specific entropy of the gas,
regardless of the change in temperature.  Similarly, ``heating''
always refers to the addition of heat to the gas, which invariably
causes an increase in its specific entropy \citep{vd05}.

\subsection{Demands of the High Incidence of Short Cooling Times}

In terms of the specific entropy, $S$, the energy equation of the gas
is
\begin{equation}
\rho T {dS \over dt} = \heat - \cool,
\label{eqn:entropy}
\end{equation}
where $\rho$ is the gas density, $T$ is its temperature, $\heat$ is
the heating rate per unit volume and $\cool$ is the power radiated per
unit volume.  This may be expressed as
\begin{equation}
{d \over dt} \ln \entind = {1 \over \theat} - {1 \over \tcool},
\label{eqn:unstableheat}
\end{equation}
where $\entind = kT / \nelec^{\gamma - 1}$ is the entropy index and
$\nelec$ is the electron density.  The cooling time is the time
required for the gas to radiate its thermal enery,
\begin{equation}
\tcool = {p \over (\gamma - 1) \cool},
\label{eqn:coolingtime}
\end{equation}
and the analogous heating time is 
\begin{equation}
\theat = {p \over (\gamma - 1) \heat}.
\end{equation}
For most heating mechanisms, the effect of excess cooling is to
increase the dominance of cooling over heating and {\it vice versa},
destabilizing the balance between heating and cooling.  If cooling
dominates, it tends to run away, so that the gas will cool
to low temperatures.  When heating dominates, the gas is driven 
toward a state where (equation \ref{eqn:unstableheat}) $\theat
\simeq t$, the age of the system, in which case the cooling time would
be $\tcool = \theat \heat / \cool \gg t$.  Thus, we should expect to
see long cooling times in systems where heating dominates.

In the absence of feedback, typical heating rates must be sufficient
to offset radiative cooling.  Otherwise cooled gas
would be deposited at rates of many hundreds of solar masses per year
in many clusters, in conflict with observations \citep{pkp03, pf06}.
On the other hand, if the balance is tipped in favor of heating, we should expect typical cooling times in excess of the
Hubble time.  This is in conflict with the observed high fraction
of short cooling times.  For example, 44\% of the HIFLUGCS sample have
central cooling times shorter than 1 Gy \citep{hmr10}.  This issue is
even sharper for lower mass systems.  Their lower temperatures suppress
thermal conduction, and there is less dynamical activity
to drive heating.  At the same time, their central cooling times are often
shorter than those in rich clusters \citep[and this volume]{svd09}.

Thus, while heating that does not involve feedback can certainly
reduce the demands on the process that prevents copious amounts of gas
from cooling to low temperatures in clusters, a process involving
feedback is required to account for the high incidence of short
cooling times.  The prime candidate is AGN heating.

\subsection{Observational Evidence for a Feedback Loop}

A true AGN feedback loop, as opposed to simple AGN heating, must
couple the AGN energy output to the reservoir of gas
fueling it.  Apparently, this applies to the hot
atmospheres of galaxies, groups, and clusters.  The trend in
Fig.~\ref{fig:heatcool} \textit{right} shows not only that radio AGN
are powerful enough to offset cooling, but that atmospheres with
larger X-ray cooling luminosities host more powerful AGN.  In other
words, the X-ray luminosity of the cooling gas measured on tens of
kiloparsec scales, and the jet power maintained by
accretion of gas near the AGN on subparsec scales, are
apparently in causal contact.  We do not understand how energy emitted from
such a small volume couples itself so efficiently to gas on such vastly
larger scales.

The prevalence of short central cooling times is another indication that hot atmospheres are maintained by feedback. Despite the high energetic output of AGN, the radiative cooling
time of the gas surrounding them remains well below one Gyr.
Fig.~\ref{fig:tcool} \textit{left} \citep{vf04} shows cooling time
profiles for the hot atmospheres of several cooling clusters.  The
cooling time profiles steadily decline from about 14
Gyr at a radius of 200 kpc to a few $10^8$ yr within the BCG.  They
are typical of atmospheres experiencing powerful
AGN activity. Instead of
shutting down cooling entirely, cooling flows are stable and
long lived.  

\begin{figure}[!ht]
\begin{center}
\includegraphics[height=0.33\textheight]{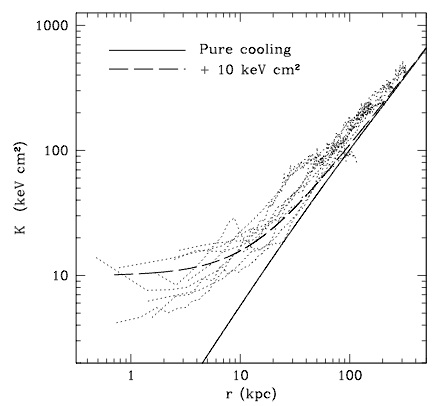} 
\includegraphics[height=0.33\textheight]{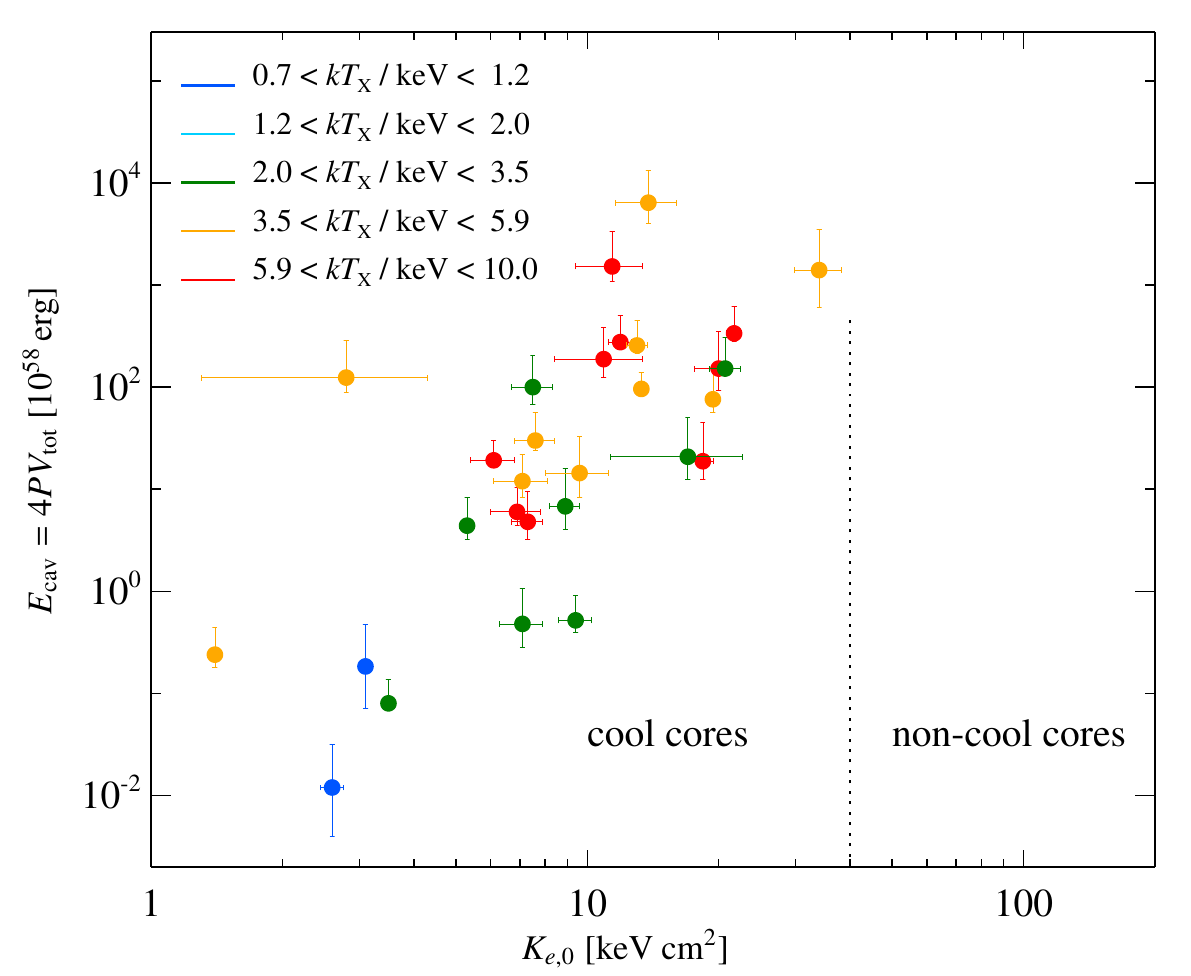}
\caption{Left panel shows atmospheric entropy profiles for a sample of
  cooling flow (cool core) clusters from \citet{vd05}.  The central
  entropy values lie above the pure cooling model.  The right panel
  shows central entropy values plotted against jet (cavity) power
  \citep{pcb11}.  The correlation shows that AGN are raising the
  central entropy. Marker colors  are indicated in the figure. }
\label{fig:pjetent}
\end{center}
\end{figure}

We shall illustrate this point using an extreme example.  
At $10^{62}$ erg, MS0735.6+7421 (MS0735) is experiencing what is arguably the most energetic AGN
heating event known \citep{mnw05}.  Yet its central cooling time is only
$5\times 10^8$ yr, which is only a few times longer than the age of the outburst.  
Assuming its AGN quenched cooling entirely and the
cooling flow shut down today,  cooling would reestablish itself and AGN
activity would resume in only several hundred Myr.  
The most powerful AGN in clusters do not dramatically raise the temperature or
lower the density of the gas in their vicinity.  Instead, they raise
the central entropy of the gas through a surprisingly gentle
``heating'' process that restores potential energy to the atmosphere that was lost by radiation
\citep{vd05, cdv09, pap10, gmb11}.  This entropy
boost is shown as a flattening of the central atmospheric
entropy profiles compared to a pure cooling atmosphere \citep{vd05}
and in the correlation between jet power and the value of the central
entropy of the host's hot atmosphere \citep{pcb11}, which are both
shown here in Fig.~\ref{fig:pjetent}.  The southwest shock of Centaurus A, which boosts
the local gas temperature by an order of magnitude, is a counter example \citep{ckh09}.  However,
such strong temperature boosts are not seen in rich clusters.  

Finally, AGN are generally active on timescales comparable to or shorter than
the central cooling timescales, a condition that is needed to prevent 
rapid cooling and puddling of cold gas in BCGs.  Fig.~\ref{fig:tcool}
\textit{right} shows central cooling time
plotted against cavity buoyancy timescales with which we estimate the age of
the AGN outbursts.  The plot shows that cavities are usually younger than the 
time it takes for the gas they are rising through to cool.  In other words,
AGN activity recurs frequently enough to prevent  runaway cooling.

\subsection{Metal-Enriched Outflows}

\begin{figure}[!ht]
\begin{center}
\includegraphics[height=0.35\textheight]{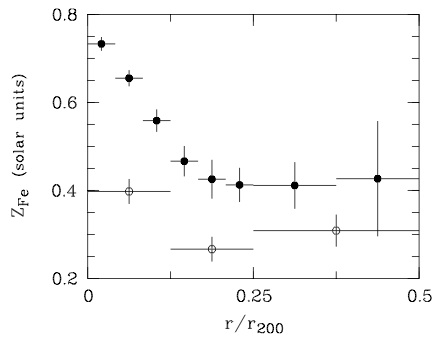}
\caption{Average iron abundance profile from \citet{del04}.  Solid
  points (open points) represent clusters with (without) a BCG.
}
\label{fig:abprof}
\end{center}
\end{figure}

Hot atmospheres surrounding clusters and galaxies are chemically enriched by stellar evolution processes
taking place in galaxies over time \citep{arb92,bft08, fbt10}. 
The gas is chemically enriched to average level of approximately
1/3 of the solar value.  However,  the metallicity approaches
and can exceed the solar value in the vicinity of
the BCG \citep{af98, dw00, dm01, del04, dwb07}.  This is shown
in Fig.~\ref{fig:abprof}  from 
\citet{del04}.  The metallicity
reaches a maximum near the solar value at the location of the BCG.
The gas has been enriched primarily by SN Ia ejecta.  The central
metallicity peak is broader than the the stellar light profile
indicating that metals are diffusing outward, well beyond the stars
that created them \citep{rcb05}.  Mixing either by mergers or AGN outflows
could produce this broadening \citep{rcb05, rbr07, rls08, dn08, gmb11}, but
AGN outflows have recently received strong observational support.
Cool, metal-rich, keV gas has been found along the
cavities and radio sources of several clusters and groups
\citep[\eg,][]{swf08, swb09, kgc09, gnd11, ogdv11, wsm10}.  The metallicity is enhanced
at levels comparable to those near the BCG, and in excess of the surrounding gas lying at the same projected radius.
This metal-enriched gas must have been launched out of the BCG along with
the radio jets and is being dispersed into the intracluster medium.

\begin{figure}[!ht]
\begin{center}
\includegraphics[height=0.35\textheight]{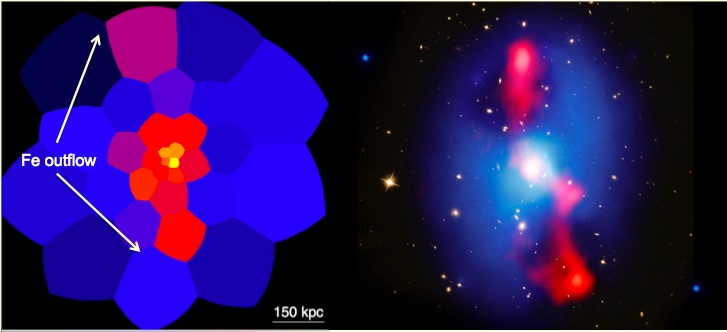}
\caption{\textit{Left:} Abundance distribution in MS0735.  Red and purple
cells indicate enhanced metal abundance compared to the ambient gas.
  \textit{Right:} X-ray (blue), radio (red) and optical images of
  MS0735 on the same scale as the abundance distribution in the left
  panel.}
\label{fig:ms07fe}
\end{center}
\end{figure}

The amount of gas being transported outward is substantial.  In Hydra
A,  several $10^7 ~\rm
M_\odot$ of iron alone has been lifted more than 120 kpc into the intracluster
medium \citep{swb09, kgc09}.  The largest known outflow
in the MS0735 cluster can be traced out to an ``iron radius" of
 $R_{\rm  Fe}\simeq 300$ kpc (Fig.~\ref{fig:ms07fe}).  Excess metallicity
 along the jet axis has been found in at least 17 clusters with cavity systems (Kirkpatrick \& McNamara in preparation).
The iron radius scales  scales surprisingly well with mechanical jet power with the form
$R_{\rm Fe} ({\rm kpc}) =  60 P_{\rm jet}^{0.43}$, where jet power is expressed in units of $10^{44}~\rm erg~s^{-1}$ \citep[; Kirkpatrick et al. 2012, in preparation]{kmc11}.  This scaling is consistent with typical radio AGN displacing metals from the BCG well
into the hot halos of clusters, as observed by De Grandi and others (Fig.~\ref{fig:abprof}). 
Kirkpatrick's scaling relation provides an 
estimate of the average jet power over several hundred Myr that 
is independent of more complicated cavity and shock front measurements.

Iron alone represents roughly one part in one thousand of the total mass flowing out with
the jets and cavities.  In Hydra A, the total outflowing gas
mass exceeds $10^{10}~\rm M_\odot$.  This implies an average outflow rate
of $\sim 100~\rm M_\odot ~yr^{-1}$ over the few $10^{8} ~\rm yr$
duration of the current outburst.  An outflow of this magnitude would
reduce the rate of mass deposition and star formation from the cooling
flow \citep{mbb08}.  A substantial fraction of the energy required to
lift the gas is likely to be dissipated when low entropy gas falls
back inwards, which could be a significant heat source in the ICM
\citep{gnd11}.  A flow of this size may be expected to deplete the
high metallicity gas at the center \citep[eg.,][]{gm10}.  However, after accounting for
replenishment of metals from ongoing stellar evolution, AGN apparently
do not erase the central iron peaks entirely, as indicated by recent
simulations of gentle feedback \citep{gmb11}.  In fact, they
probably prevent excessive central metal enrichment and subsequent
overcooling.

A radio AGN's ability to lift volume-filling gas depends on its 
mechanical power and not its luminosity.  The iron radius is 
 thus limited by the accretion rate onto the black hole in Eddington units.
As the SMBH grows and the accretion rate falls below a few percent
of the Eddington rate, the AGN is expected to transition from a
quasar to a mechanically-dominated radio jet (see \citealt{nm08} for a review, and below).  The transition from a quasar to a radio galaxy would be accompanied by a dramatic rise in the work done by the jet on the volume-filling gas, and a hot outflow would ensue.  
Thus, the dispersal of metals throughout  cluster atmospheres by AGN 
probably occurred primarily during the later stages of
cluster development following the putative quasar phase of the BCG.

\section{How AGN Outbursts Heat Hot Atmospheres}

It remains unclear which processes are
communicating the energy of AGN outbursts to the hot atmospheres that
host them, or how outburst energy is dissipated in the gas.  Even if
power flows exclusively via jets into the radio lobes, there are
several channels for transfering the power to the ICM.  In the basic
model outlined by \citet{s74}, jets inflate radio lobes which do work
on the gas as they expand.  Thus, jet power is divided between the
lobes and the ICM.  A rapidly inflating lobe can drive shocks into the
gas, producing edges in X-ray surface brightness that have been
observed in a number of objects \citep[\eg,][]{swa02, jfv02, kvf03,
  mnw05, mnj06}.  However, the typical division between
internal energy in the lobes and mechanical work done on the ICM has
yet to be assessed \citep[estimates exist for some cases,
  \eg,][]{nmw05, fnh05, rfg11}.  This energy division also evolves
with time.

While theory suggests that the lobes should be unstable to rapid
disruption, observations indicate that they can survive for long periods of time
\citep[\eg,][]{fsa11}.  Disruption can be delayed if the instability
is suppressed by viscosity \citep{rmf05}, turbulent diffusion
\citep{sb08}, magnetic stresses \citep{reb07, odj09}, or favorable
dynamics \citep{ps06}.  Regardless of the details, ultimately the lobe
plasma is likely to mix with the ICM, depositing cosmic rays and
magnetic energy there.  In hydrodynamic jet simulations
\citep[\eg,][]{bk02, hby06, deyoung10} heat is transfered to the ICM when lobe
gas is mixed with it, but this may be unrealistic.  Energy exchange
could be much slower if the primary forms of lobe energy that are
mixed with the ICM are cosmic rays and magnetic fields
\citep[\eg,][]{bm88, mg10}.

Cosmic rays should also diffuse out of radio lobes into the
surrounding ICM \citep{bm88, sps08}.  In addition to slow collisional
heating, streaming cosmic rays can excite plasma waves that are likely
to end up dissipating and acting as another source of heat
\citep{epm11}.  Indeed, \citet{go08} argue that cosmic ray heating
powered by the central AGN, coupled with a significant level of
thermal conduction can stabilize the cooling gas.  If lobes remain
intact as they rise buoyantly, they will maintain approximate
pressure balance with their surroundings.  Their pressure
decreases with time causing a decrease in enthalpy that
releases energy into the ICM, mostly as kinetic energy in the flow
around rising ``bubbles'' \citep{mn07}.

The larger scale gas flow created by an expanding lobe can be made turbulent by
interaction with small cavities left by earlier generations of AGN
 \citep{hc05}, or by interaction with dense clouds
\citep[\eg,][]{ef03, sc03, scr11}.  Uplift of low entropy gas with
rising cavities and the fallback that is likely to follow
\citep{cbk01, gnd11} and flow around buoyant bubbles
can also generate turbulence in the ICM \citep{sfs11}.  This turbulence will be
dissipated, providing yet another channel for dispersing outburst
energy into hot atmospheres \citep[\eg,][]{dc05}.

While it seems likely that enough energy is being deposited by AGN
outbursts to prevent hot atmospheres from cooling to low
temperatures, two key issues remain to be resolved: whether enough of
the outburst energy is dissipated to prevent the gas from cooling and
how that energy is distributed throughout the gas.  In the remainder of this
section we consider some aspects of heating and cooling in more detail.

\subsection{Adiabatic Uplift}

The decrease in specific entropy caused by radiative losses is
balanced most effectively by dissipative processes that add heat and
increase entropy.  Beyond energy deposition, dissipation is needed to
counteract the heat loss due to radiative cooling.  To emphasize this
point, we consider what would happen if there was no significant
heating in an AGN outburst.  

\begin{figure}[!ht]
\centerline{\includegraphics[width=0.6\textwidth,angle=270]{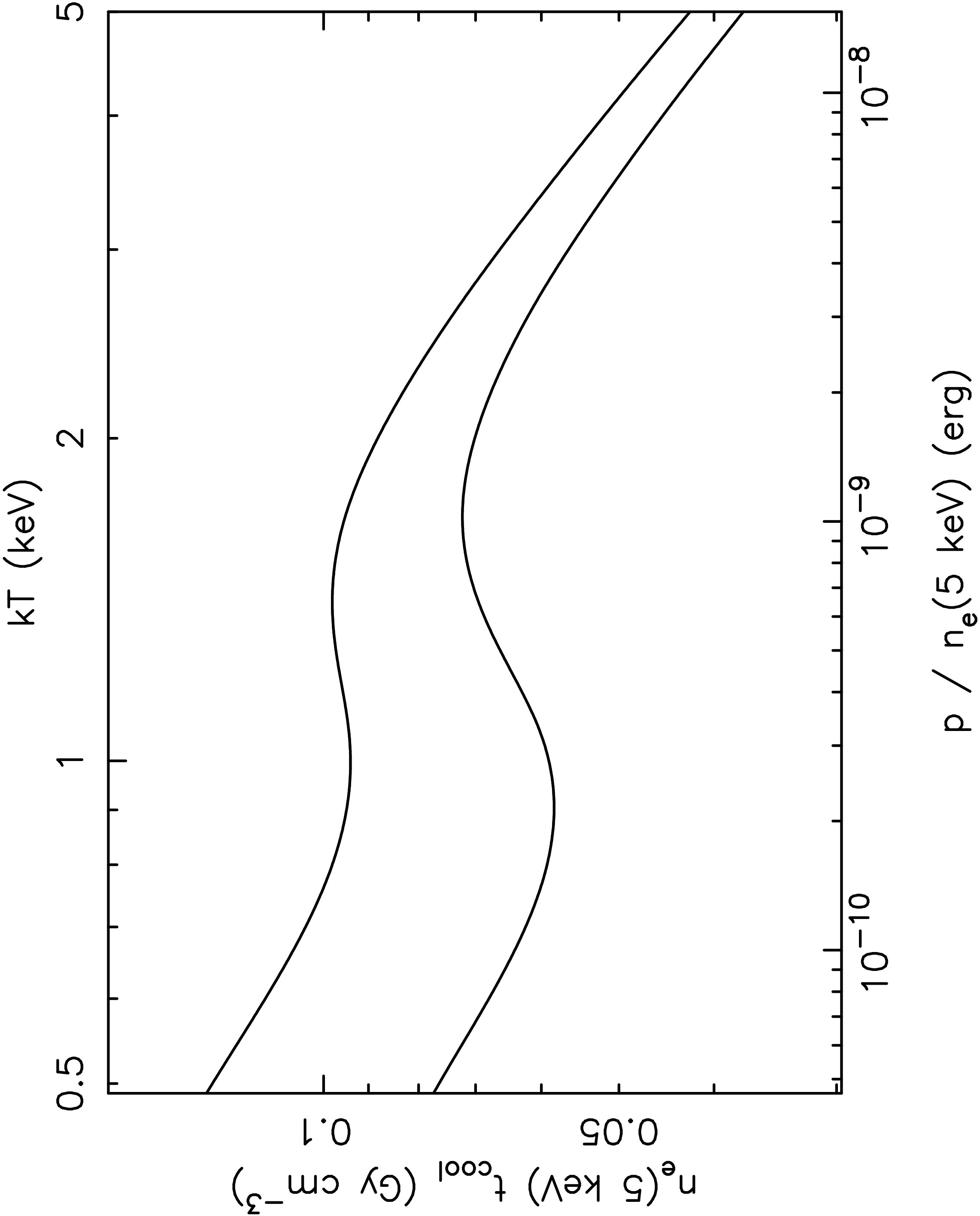}}
\caption{Cooling time \vs{} pressure under adiabatic expansion.
  Cooling time (Gyr) multiplied by the electron density ($\rm
  cm^{-3}$) at 5 keV is plotted against the pressure divided by the
  electron density at 5 keV for gas with solar (lower curve) and half
  solar abundances (upper curve), under an isentropic pressure change.
  The upper scale shows the temperature in keV.  The cooling function
  used is from the APEC model \citep{sbl01}.  Below $\sim 2$ keV, the
  increase in the cooling function due to line cooling shortens
  cooling times significantly.  Particularly for $kT \lesssim 3$ keV,
  typical of the temperature in cool cores, the cooling time is quite
  insensitive to adiabatic pressure changes.}
\label{fig:adtcool}
\end{figure}

The state of a simple gas is completely determined by its entropy and
pressure, so that we only need to consider the effect of the change in
gas pressure.  Expressed in terms of the cooling function, $\Lambda
(T)$, the cooling time (equation \ref{eqn:coolingtime}) is
\begin{equation}
\tcool = {p \over (\gamma - 1) \nelec \nh \Lambda(T)},
\end{equation}
where $\nelec$ and $\nh$ are the electron density, and proton density
of the gas, respectively.  For isentropic gas, $T \propto p^{(\gamma -
  1) / \gamma}$ and $\nelec \propto p^{1 / \gamma}$, indicating the
pressure dependence of the cooling time.  Fig.~\ref{fig:adtcool} shows
how the cooling time varies under an adiabatic pressure change for gas
with solar and half solar abundances.  Adiabatic uplift would decrease
the pressure.  The increase in cooling time as gas is decompressed by
2.5 orders of magnitude, from an initial temperature of $kT = 5$ keV
(hotter than the central gas in most cool cores with short central
cooling times) is just over a factor of 2 for solar abundances
(typical of the central gas in these systems).  For half solar
abundances the increase is a little over a factor of 3.  For observed
abundances, in the temperature range of interest, the cooling time is
quite insensitive to the pressure.  Thus, lifting gas from the center
of the cool core to regions of much lower pressure delays cooling to
low temperatures by no more than a factor of $\simeq 3$ and,
typically, considerably less.  For gas with cooling times $<1$ Gyr,
adiabatic uplift is ineffective as a way to prevent most of the gas
from cooling to low temperatures.

\subsection{Lobe Size and the Spectrum of AGN Outbursts}

The timing and power of AGN outbursts, \ie, their
power spectra \citep[\cf][]{nb05} are important issues 
affecting the properties of radio lobes and their
effect on the ICM.  When an outburst deposits energy,
$E_{\rm tot}$, into a volume, $V$, the pressure increase is $\sim
(\gamma - 1) E_{\rm tot} / V$.  This makes the fractional pressure
increase large if the total thermal energy within $V$, \ie, $\int_V
p/(\gamma - 1) \, dV$, is smaller than $E_{\rm tot}$.  With a large
fractional pressure increase, the gas in $V$ would expand rapidly,
driving shocks into the surrounding gas.  Thus, the smallest volume
that can be affected by an AGN outburst contains a thermal energy
comparable to the energy deposited by the outburst, placing a lower
limit on the sizes of radio lobes and cavities.
Conversely, a new outburst injecting energy into a existing radio lobe
can have very little effect on the surrounding gas unless the total
energy injected is comparable to its existing internal energy.  Sizes
and shapes of lobes are also affected by the detailed properties of
jets and their interaction with the environment.  Lobe size clearly
affects where the outburst energy is deposited in the environs of an
AGN.  That is, the spatial distribution of the outburst energy is an
important property for the feedback process.

The average power is another important factor.  Even
if the jet power is constant, $pV$ generally increases faster than
linearly with the radius, making the speed at which a lobe advances
decrease with time in the absence of other effects.  When the
expansion is subsonic, buoyant lift may detach a lobe from its jet and
turbulent motions in the surrounding ICM may toss the lobe around. 
Simulations have shown that turbulence in the ICM
driven by the continuing infall of subhalos can have a marked effect
on the structure of radio lobes \citep{brh05, hby06}.  Based on a
small set of simulations, \citet{mhb10} found that the size of the
region affected by an outburst, its ``radius of influence,'' $R$, is
determined by outburst power, $P_{\rm j}$, with $R \propto P_{\rm
 j}^{1/3}$.  In these simulations, ICM ``weather'' (turbulence) also
leads to relatively isotropic deposition of the AGN energy.  While
disruption and diffusion is the likely fate of all lobes, a number of
systems, such as MS0735.6+7421 \citep{mnw05} and NGC~5813
\citep{rfg11}, show considerable large-scale symmetry that has
survived through multiple outburst cycles.

\begin{figure}[!ht]
\centerline{\includegraphics[width=0.6\textwidth]{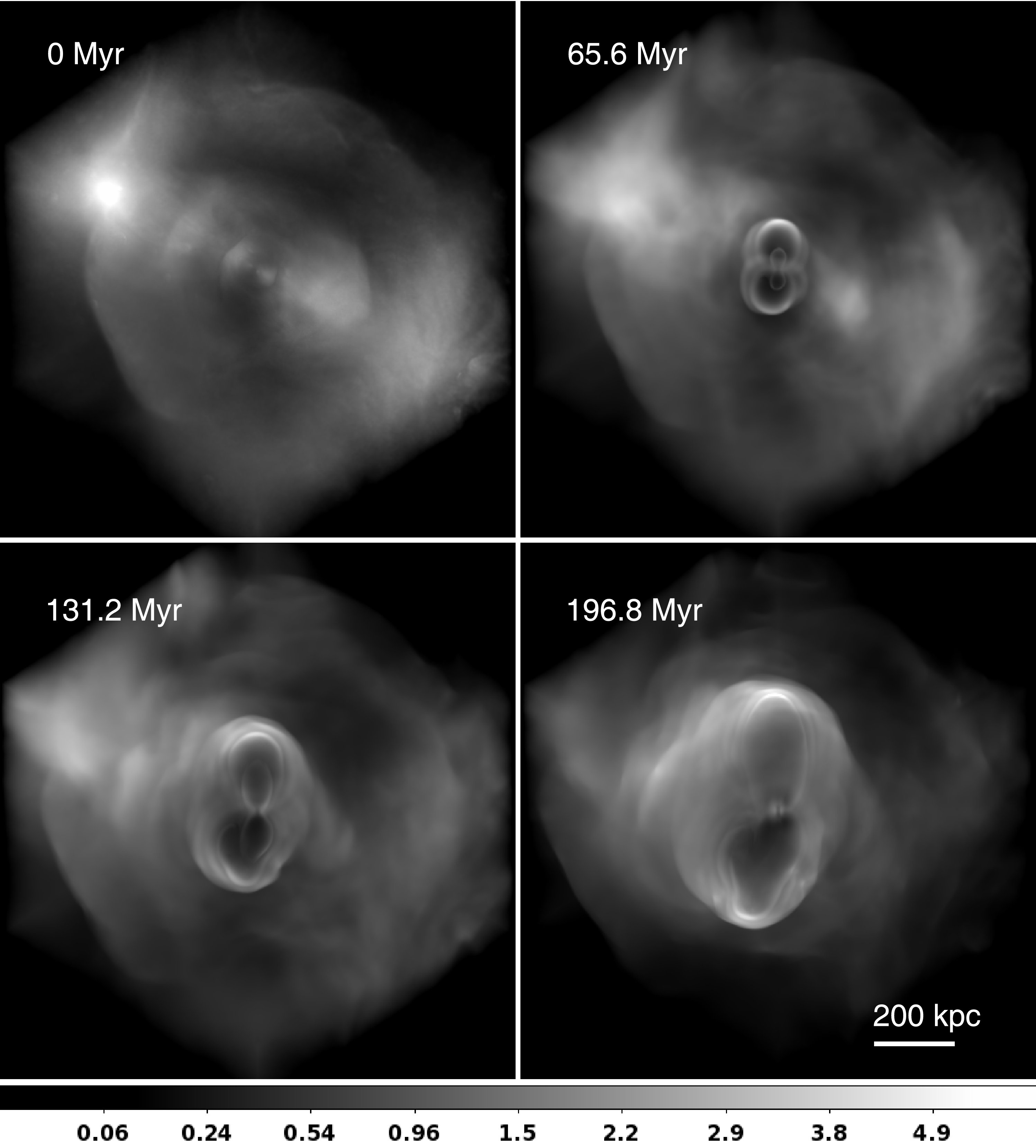}}
\caption{Shocks and cavities for a simulated intermittent AGN at a
  cluster center.  These are synthetic broadband X-ray images, divided
  by a beta model, at various times, made from the MHD simulations of
  Mendygral \etal{} (2012).  Their simulations represent a pair of
  intermittent, opposed AGN jets, with 50\% duty cycle, at the center
  of a $\simeq 1.6$ keV cluster extracted from a cosmological
  simulation.  In contrast to the highly elongated lobes and shocks
  produced by jets with constant power, these lobes and cavities
  resemble the moderately elliptical systems found in reality. }
\label{fig:mendygral}
\end{figure}

Other significant properties affected by outburst history are the
shapes and relative positions of the cocoon shocks and lobes.  While a
steady, continuous jet produces highly elongated shock fronts
\citep{vr07}, simulations \citep{oj10} show that intermittent
outbursts can make the fronts and lobes considerably more spherical,
as observed.  This is illustrated in Fig.~\ref{fig:mendygral}, which
shows synthetic X-ray images from MHD simulations of a pair of jets
with a 50\% duty cycle at the center of a realistic cluster, from
Mendygral, Jones, \& Dolag (2012, submitted).

\subsection{Weak Shock and Sound Heating}

Outburst energy can appear in the gas as thermal, kinetic, or potential energy.
Much of its impact is transient.  For example, a passing
weak shock compresses and heats the gas, accelerating it to a moderate
fraction of the sound speed.  However, when the shock has passed the
gas returns to rest, its pressure settles to slightly less than the
preshock pressure because the gas has moved outward where
gravity is weaker.  A weak shock raises the entropy only slightly,
 so the reduction in pressure leaves the gas cooler than before
the shock passed.  Most of the residual energy gain from the shock remains as
gravitational potential energy.  All this happens in a time that is
small compared to the cooling time, so that the transient effects of a
single weak shock are largely negligible.  Using the example of the
Hydra A cluster, \citet{dnm01} and \citet{njf07} argued that weak shock heating fails to
compensate for radiative losses in cool cores.

The primary effect of radiation is entropy reduction.  The entropy
increase in a single weak shock, $\Delta S$, provides an effective
heat input of $\Delta Q \simeq T \Delta S = E \Delta \ln \entind$,
where $\entind$ is the entropy index (see below equation
\ref{eqn:unstableheat}), $E$ is the specific energy and
$\Delta\ln\entind$ is the change of $\ln\entind$ in the shock.  The
fractional heat input, $\Delta Q/E$, in a single weak shock is small,
but the cumulative effect of multiple shocks can be significant.  Two
nearby, well-observed systems, M87 \citep{fjc07} and NGC~5813
\citep{rfg11}, have experienced multiple shocks. The time interval between them
is short compared to the cooling time.  If repeated weak shocking is sustained 
over long periods of time,  it can comfortably compensate for the
heat radiated in the inner regions of both M87 and NGC~5813.  Because
shock strength generally decreases with radius and the entropy
increase is cubic in the shock strength, weak shock heating is most
effective at small radii.  Thus, it appears likely that the process
responsible for preventing gas from cooling to low temperatures close
to the Bondi radius is weak shock heating in at least these two
systems.  Given the importance of gas properties at the Bondi radius
to feeding of the AGN (see section \ref{sec:bondi}), this gives weak
shock heating a potentially critical role in the feedback process.
Other heating processes are required and are more effective on larger
scales.

Sound waves in the Perseus cluster ICM created by repeated outbursts
from NGC~1275 could be a significant heat source \citep{fsa03, frt05}.
Thermal conduction and viscosity cause the sound to dissipate,
generating significant heat if the transport coefficients are close to
their magnetic field free values.  \citet{rbb04} showed that heating
by sound dissipation might also be able to offset cooling in the long
term.  Whereas weak shock heating depends on the cube of the shock
strength (pressure amplitude), sound heating is proportional to the
square of the wave amplitude, making it more effective on larger
scales.  Unfortunately, while the shock heating rate is largely
independent of the uncertain transport coefficients, sound heating
does depend on them (section \ref{sec:transport}).

\subsection{Isotropic Jet Heating}

\begin{figure}[!ht]
\begin{center}
\includegraphics[height=0.35\textheight]{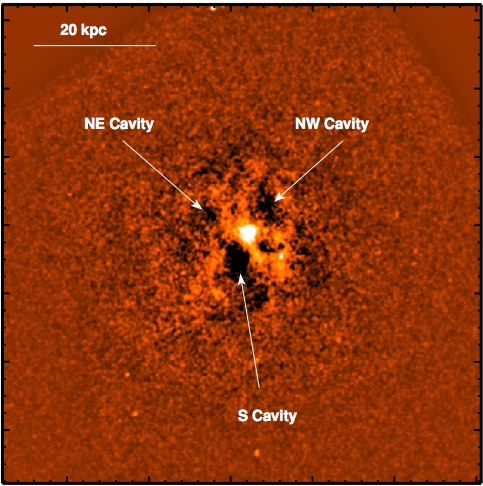}
\includegraphics[height=0.35\textheight]{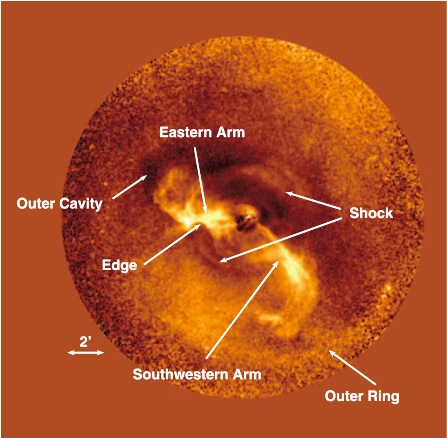}
\includegraphics[height=0.3\textheight]{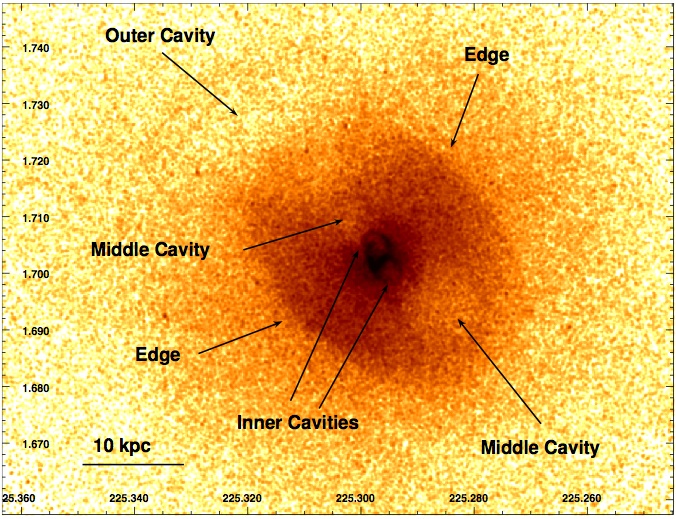}
\includegraphics[height=0.3\textheight]{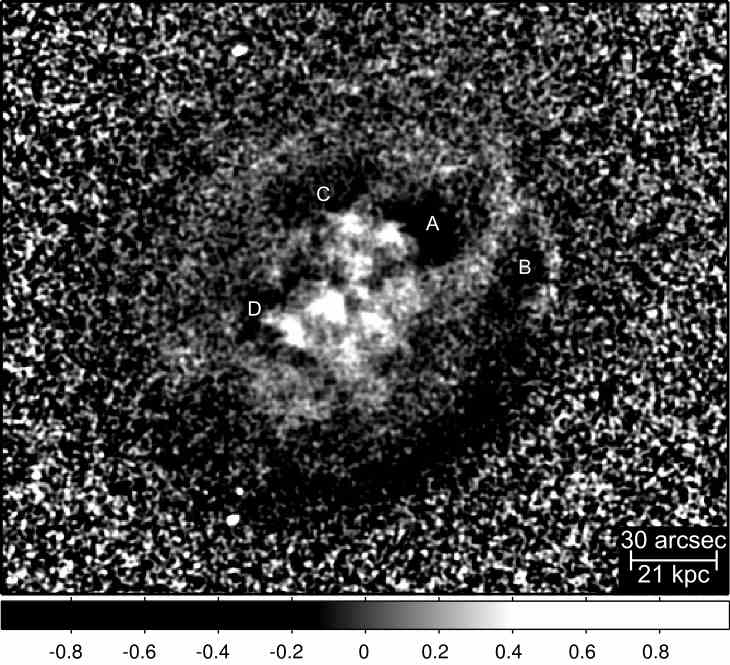}
\caption{Examples of systems showing multiple cavities and shocks.
  \textit{Upper left:} Unsharp mask 0.3 -- 2 keV image of the NGC~5044
  group showing multiple cavities \citep{djf09}.  \textit{Upper
    right:} Region around M87 in the Virgo cluster, showing deviations
  from the azimuthally averaged surface brightness in the energy band
  0.5 -- 2.5 keV \citep{fjc07}. \textit{Lower left:} 0.3 -- 2 keV
  image of the NGC~5813 group, showing multiple shock fronts and
  cavities \citep{rfg11}.  \textit{Lower right:} Unsharp mask image of
  2A0335+096 showing multiple cavities and sound
  waves\citep{sft09}. } \label{fig:cav}
\end{center}
\end{figure}

\begin{figure}[!ht]
\begin{center}
\includegraphics[height=0.45\textheight]{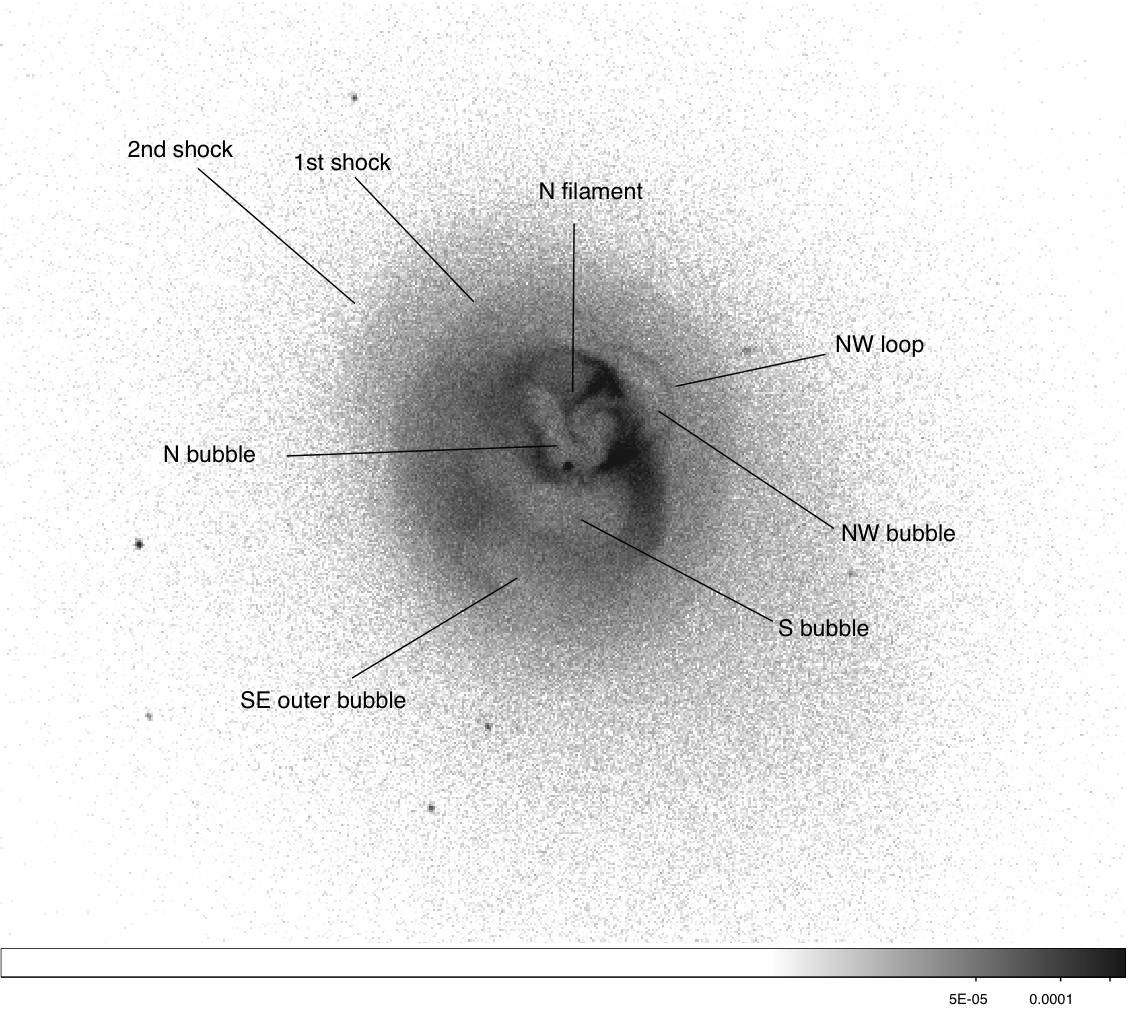}
\caption{Deep, 650 thousand second exposure of Abell 2052 taken with
the Chandra X-ray observatory showing multiple cavities and 
weak shock fronts in the inner 60 kpc of the cluster surrounding the central radio source \citep{brc11}. } \label{fig:a2052}
\end{center}
\end{figure}

Radio synchrotron jets, such as Cygnus A (Fig.~\ref{fig:cyga}), give
the impression that the jet's energy flux is confined to a narrow
tube.  While this may be true initially as the jets are launched,
X-ray and low frequency radio images show that jet power is quickly
dissipated over a large solid angle.  In canonical systems like M87
\citep{fjc07} and Perseus \citep{fsa11}, as well as lesser known
systems such as NGC 5044 \citep{djf09} and 2A0335+096 \citep{sft09},
energy is being deposited by multiple cavities, weak shocks, and sound
waves that are dissipating energy over much of the volume of the inner
atmosphere (Fig.~\ref{fig:cav}). 
%% Only a percent or so of the energy of a sound wave actually heats
%% the gas.  However, o
Observations of the Perseus cluster \citep{fsa03, frt05}, M87
\citep{fjc07}, Abell 2052 \citep{brc11} (Fig.~\ref{fig:a2052}),
Centaurus \citep{fst05, sf08} have shown that their AGN generate sound
waves of sufficient intensity to heat the central volume of gas.
Furthermore, cavities are emerging in multiple directions from the nuclei
of the elliptical galaxy NGC 5044 and the BGG in 2A0335+096 (Fig.~\ref{fig:cav}).
This may indicate that jets are launched by a series of
randomly-oriented accretion disks \citep{kp07}, jet precession
\citep{dfs06, grs11} and/or the larger-scale cavities are being blown
around by atmospheric pressure gradients \citep{brh05, hby06, mhb10}.

Even in cases where jet heating is highly anisotropic, following an outburst the
lowest entropy gas that remains, which is the most in need of heating,
falls towards the cluster center.  Since the free-fall time is short
compared to the cooling time, there is more than enough time for this
gas to fall in close to the AGN where it can be heated most
effectively in subsequent outbursts.  Such circulation is observed in simulations of unidirectional jets (Peter Mendygral, private communicaton). The need for isotropic heating
by apparently anisotropic jets has been cited as a problem by many
researchers.  In our view, this issue is not as serious as is commonly
perceived.

\subsection{Transport} \label{sec:transport}

Poor understanding of the transport properties of hot atmospheres has impeded
progress.  In the absence of magnetic fields,
transport properties are governed by Coulomb collisions.  The
Coulomb mean free paths for electrons is $\simeq 0.26 (kT)^2
\nelec^{-1}$ pc and for protons $\simeq 0.37 (kT)^2 \nelec^{-1}$
pc.  Here, $kT$ is given in keV and $\nelec$ in $\rm cm^{-3}$.  Thermal conduction
would certainly play a significant role in helping to
prevent gas from cooling \citep[\eg,][]{zn03, vf04} and the viscosity
would be dynamically significant \citep[\eg,][]{fsc03}.  However, the
ratio of the Larmor radius to the mean free path for thermal electrons
is $r_{\rm L,e} / \lambda_{\rm ee} \simeq 1.3\times10^{-10} \nelec
(kT)^{-3/2} \bmug^{-1}$ and for thermal protons it is $r_{\rm L,p} /
\lambda_{\rm pp} \simeq 4\times10^{-9} \nelec (kT)^{-3/2} \bmug^{-1}$,
where $\bmug$ is the magnetic field strength in $\mu\rm G$
\citep[typically $\bmug\sim$ few in the ICM, \eg,][]{ct02, bgf11}.  The particles are 
coupled tightly to the magnetic field lines
and transport is extremely anisotropic.  In the presence of a weak
magnetic field tangled by isotropic turbulence, the
effective thermal conductivity is expected to be reduced by a factor of $\simeq 3$ \citep{nm01},
presumably with a similar effect on the viscosity.  However, the
true situation may be far more complex.

\subsubsection{Magnetothermal Instability}

Anisotropic conduction introduces important new effects by making
radial temperature gradients unstable.  With a negative radial
temperature gradient, an atmosphere is subject to the magnetothermal
instability \citep[MTI;][]{b00, ps05}.  In essence, conduction tends
to keep gas on the same field line isothermal.  When a parcel of
gas lying on a horizontal field line moves outward while maintaining its
temperature at the same value as the bulk of the gas on the same field
line, it will be hotter than its new surroundings.  Then buoyancy makes it
rise further, causing instability.  Simulations show that MTI drives turbulence
and orients the magnetic field preferentially in the radial direction,
helping to promote radial heat flux \citep{psl08}.

Negative temperature gradients occur in the outer atmospheres of clusters
and throughout non-cool core clusters \citep[\eg,][]{vmm05}.  However,
the temperature gradient is generally positive in cores
with short central cooling times, so that MTI is not important
for AGN feedback in massive clusters.  Less massive hot 
atmospheres generally have short central cooling times and often do
not show temperature drops in the core.  These systems may be affected by
MTI. However, the steep temperature dependence of the thermal conductivity
\citep[$\kappa \sim T^2$,][]{b65} limits the role of conduction at
lower temperatures.  Since the power available to drive the turbulence
cannot exceed the conductive heat flux, MTI will be less vigorous in
cooler systems.

\subsubsection{Heat Flux Driven Buoyancy Instability}

With a positive radial temperature gradient, highly anisotropic
conduction makes an atmosphere subject to the heat flux driven
buoyancy instability \citep[HBI,][]{q08}.  When the magnetic field is initially parallel to the temperature
gradient, there is a steady heat flux, ${\bf h}$, along the field
lines.  Inclined perturbations then kink the field lines, producing
small scale variations in $\nabla\cdot{\bf h}$, so that heat is
deposited in some regions and removed from others.  The resultant
heating and cooling can then drive buoyant motions that amplify the
perturbations.  Like MTI, HBI drives turbulence, but it tends to
orient field lines perpendicular to the temperature gradient, strongly
suppressing thermal conduction \citep{brb09, pqs09}.  Under the right
circumstances, turbulence driven by HBI and MTI may enhance diffusive
transport in the ICM \citep{scq09}.

Cool core clusters generally have positive radial temperature
gradients in their cores, and HBI may suppress thermal conduction
in them.  However, turbulence driven by other processes, including
moving subhalos, continuing minor mergers, and AGN activity, can
overwhelm the relatively weak effects of HBI, isotropizing the
magnetic field and maintaining a relatively high level of thermal
conduction. The power available to drive HBI turbulence is
  also limited by the total conductive heat flux \citep{pqs10}.
\citet{ro11} argue further that turbulence driven by moving subhalos
in cluster cores is sufficient to make turbulent heat diffusion
dominate over conduction there.

\subsubsection{Plasma Effects}

Typically, the magnetic pressure in the ICM is only a small fraction,
$\sim 1\%$, of the gas pressure \citep{ct02, bgf11}, so that the
magnetic field is dynamically insignificant.  The field has little
influence on fluid motions while it is dragged along with it.  Thus,
gas motions tend to produce changes in the magnetic field strength.
In the absence of particle-particle collisions, adiabatic invariance
of the magnetic moment would then change the particle velocity
dispersion perpendicular to the field, making the pressure
anisotropic.  Collisions counteract this effect, but the collisonal
relaxation time for protons in the ICM, $\tau_{\rm pp} \simeq 700
(kT)^{3/2} \nelec^{-1}$ yr, is long enough that fluid motions can
create significant levels of pressure anisotropy.

\citet{sck05} argued that when the field is weak, pressure anisotropy
due to fluid motions leads to fire hose or mirror instabilities that rapidly
strengthen the field until the
magnetic pressure is comparable to ${\rm Re}^{-1/2} p$, where $\rm Re$
is the Reynolds number.  \citet{scr10} have identified another
plasma instability, gyrothermal instability, that is triggered by heat
flux along the magnetic field.  They argue that gyrothermal instability could limit the
parallel heat flux.  More generally, they make a strong case that
transport properties of the plasma cannot be modeled simply in terms
of a conductivity and viscosity like unmagnetized plasmas.

\citet{ksc11} argue that these processes can provide a stable local
heating mechanism.  Assuming that turbulence can maintain the pressure
anisotropy close to the margin of the firehose or mirror
instabilities, they equate it to the pressure anisotropy resulting
from competition between a steadily changing magnetic field and
collisional relaxation. This provides a dissipation rate expressed in
terms of thermodynamic properties of the fluid and the magnetic field
strength that can stably balance radiative cooling.  However, this
mechanism requires an external source to drive sufficient turbulence
to maintain the pressure anisotropy.  The primary dissipation channel must also be that
associated with the pressure anisotropy, which is generally slow
compared to the turnover time of the largest eddies in turbulent systems.

%\textit{Do we want
 % to delete this --- or reduce it to just the first sentence?  The
 % more I think about it, the more bogus it looks.  Ignoring other
  %issues, just consider stirring their plasma.  If you stir hard
  %enough to maintain their level of turbulence, then the stirring
  %power must, at least, match the dissipation rate.  If the stirring
  %power exceeds the dissipation rate, turbulence would build up with
  %time.  To avoid things getting silly, they would argue that the
  %extra power builds B field until the stirring power matches the
  %dissipation rate.  OK, so let's suppose that things adjust to make
  %the level of turbulence balance the cooling rate locally.  Now
  %introduce some density variations.  According to their reasoning,
  %extra power has to be sucked out of the turbulence locally to
  %balance cooling in the regions where the density is higher.  If you
  %buy the argument so far, then that would imply a local reduction in
  %the B field, but, since the dissipation rate $\sim B^4$, the heating
  %rate would dive - they simply ignore the possibility that B changes
  %at this point in their reasoning.  Alternatively, they have to find
 % some mechanism to increase the local stirring power input.}

Simulations show that turbulence tends to create an intermittent
magnetic field distribution \citep[\eg,][]{jpr11}.  Turbulent motion
tends to expel magnetic field from much of the ICM, leaving a weak
field occupying most of the volume and strong fields confined to small
regions.  Little work has been done on the impact of intermittent
magnetic field on transport.  Nevertheless,  the field distribution and the
relative sizes of the collisional mean free paths and magnetic field
structures are likely to affect transport properties.  At present, we
can only conclude that transport phenomena in the ICM may be complex.
Ignorance may yet prove to be the only basis for
assuming that transport in the ICM is any simpler than in the solar
wind.

\section{The Star Formation-Cooling Time-Entropy Threshold
} \label{sec:sftc} 

A link between cooling hot atmospheres and the cold interstellar medium was established
thirty years ago with the discovery of bright  H$\alpha$
emission in BCGs located
in cooling flow clusters \citep[\eg, Fig.~\ref{fig:perseus} \textit{right}]{heckman81, hcw85,
  hbv89}.  Most of these BCGs are forming stars
at rates of several  to several tens of solar masses per year
\citep{jfn87, mo89, obp08}. Their rates dramatically exceed those in
normal giant elliptical galaxies and most nearby spirals. Therefore, they
cannot be maintained by stellar mass loss. (See \citealt{vd11} for a more nuanced
view.)  Nevertheless, their star formation rates are only a few percent
or less of the rates expected from pure radiative cooling. 

The gulf between X-ray cooling rates and star formation rates was for
decades the most serious challenge to cooling flow models.  
This situation changed dramatically when X-ray spectra of cooling flows
obtained with the \xmm{} observatory failed to detect emission
features from Fe XVII and other species at the expected levels \citep{pkp03,pf06,sff10}.  
Cooling gas is found with a range of temperatures,
but  there is less emission from low temperature gas than classical cooling
flow models predict \citep{pkp03,ktp04}.  Furthermore, the emission measure distribution of the
cooling gas is apparently inconsistent with predictions of simple
cooling and heating models \citep{pkp03}.  Despite this problem, once AGN heating is accounted
for, the observed star formation rates lie within a factor of a few of the
reduced (net) cooling rates \citep{rmn06, bvk07, obp08, hmr10, dbw10, hmd10}.
The storied discrepancy between X-ray cooling rates and star formation rates is
no longer a serious issue.

\begin{figure}[!ht]
\begin{center}
\includegraphics[height=0.345\textheight]{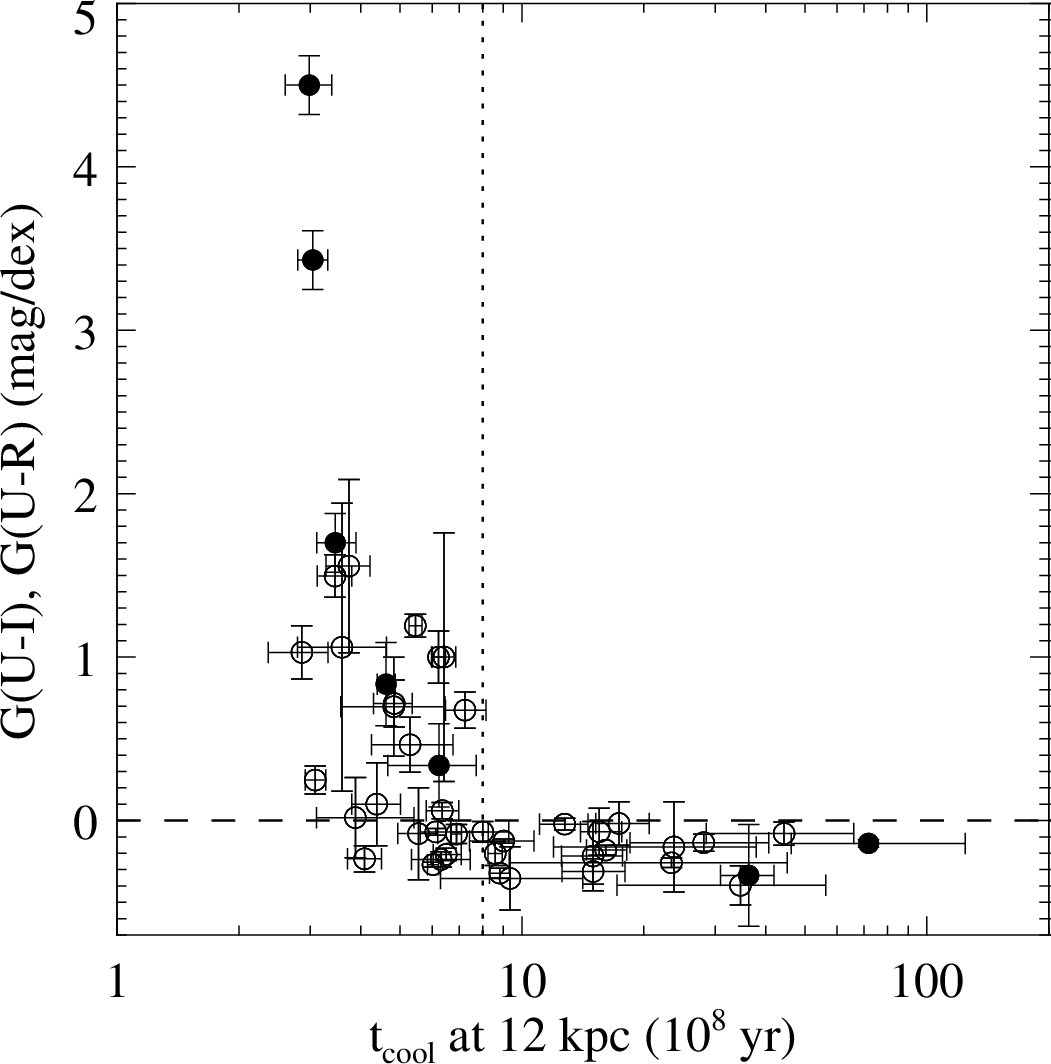}
\includegraphics[height=0.35\textheight]{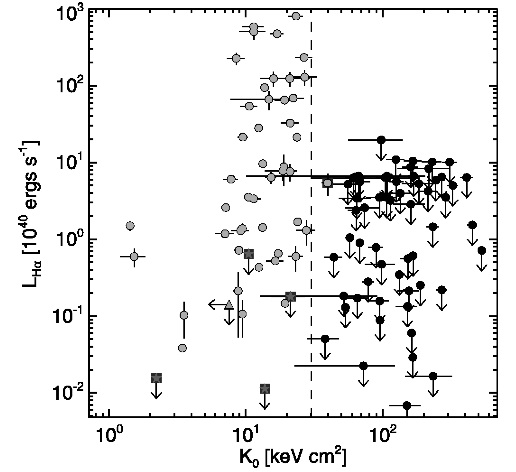}
\caption{\textit{Left:} Color \vs{} cooling time at 12 kpc
  \citep{rmn08}.  \textit{Right:} H$\alpha$ luminosity \vs{} central
  entropy \citep{cdv08}.  Young stars and line emission nebulae are
  only seen in BCG's with short central cooling times, or,
  equivalently, low central entropies.}
\label{fig:sfthresh}
\end{center}
\end{figure}

%[reference Ming Sun]

The star formation that is observed is almost certainly being fueled
by cooling flows.
The relationship between short X-ray cooling times and star formation
has recently been resolved by \chandra{} into a sharp
threshold \citep{rmn08, cdv08}.  Central host galaxy color, which is sensitive to star formation,
 is plotted against the cooling time within the BCG
in Fig.~\ref{fig:sfthresh} \textit{left}.  This diagram shows that the colors of BCGs centered in
atmospheres with cooling times exceeding $t_{\rm crit} \sim 5 \times
10^8$ yr are red and thus are not experiencing significant levels of star
formation.  However, when the central cooling time falls below $t_{\rm crit}$,
most BCGs become bluer than normal and thus are experiencing substantial levels of star
formation.  
This cooling time threshold corresponds to an entropy threshold of
about $30 ~\rm keV~cm^2$,  where \citet{cdv08} found the onset of
H$\alpha$ emission and radio activity.  
The tendency for
BCGs experiencing star formation to reside in clusters lying 
above the cluster $L_{\rm x}-T_{\rm x}$ relation 
\citep{bhb08} is a consequence of the star formation threshold.

The origin of this threshold is unclear.  \citet{vcd08} and
\citet{v11} have suggested it appears when gas becomes thermally
unstable in the competition between radiative cooling and thermal
conduction.  \citet{smq11} suggested that for thermal instability to
grow and for cooling gas to condense into stars, its growth time must
also be shorter than about 10 free-fall times.  Other clues to its
origin may be gleaned from the red BCGs with short cooling times.
These systems are usually offset from the X-ray centroid and may not
be accreting gas, or perhaps they are quenched by AGN feedback
\citep{rmn08}.  Whatever its specific origin may be, the threshold
links cooling atmospheres to star formation.

Some have suggested that the cold gas and star formation in
BCGs are merger debris or stripped material from donor galaxies
\citep[\eg, see][]{smg89}.  While this may be true in some cases, it
would only account for the prevalence of star formation in cooling
cores and the star formation threshold with great difficulty.  Galaxies located
in the cores of clusters are
largely devoid of cold gas and star formation \citep{bvk07}.
BCGs in cooling flows are the only nearby population of
ellipticals experiencing significant levels of star formation.  In the
extreme, they harbor $>10^{10} M_\odot$ of molecular gas \citep{e01,sc03}.
No known population of galaxies is able to donate molecular gas at this level.  The
most gas-rich spiral galaxies in nearby clusters contain $\sim
10^9M_\odot$ or less of molecular gas, and such a galaxy plunging 
through a cluster would release most of this gas into the ICM before it arrived at the BCG
\citep{kmc11}.   Gravitational
torques may play a role in triggering star formation
\citep{wes09}, but galaxy interactions are unable to explain the
aggregate star formation and gas properties of BCGs.

\section{The Radio AGN Itself} \label{sec:method}

%A poorly understood aspect of AGN feedback is the AGN
%itself.  
The canonical picture of a radio loud AGN involves a
$\sim 10^8~M_\odot$ black hole with a horizon radius of $3\times
10^{13}~\rm cm$, a thin accretion disk spanning radii of $\sim 1$ --
$30\times 10^{14}~\rm cm$, and a dusty torus with an inner radius of
$\sim 10^{17}~\rm cm$ \citep{up95}.  A black hole embedded in $ 1$
keV gas at the cluster center has a Bondi radius of about
$10^{19}$ cm, or a few parsecs.   It is usually assumed that 
AGN are powered by accretion onto the
SMBH \citep{lb69,bbr84}. But whether energy is released through hot or
cold accretion, how power output is related to the structure of
the AGN and the spin of the black hole, and how AGN regulate atmospheric
cooling on scales of $10^{24}$ cm  are poorly
understood.  The feedback cycle involves coordinated processes operating over 10
decades in scale, presenting both fundamental and numerical challenges
\citep[e.g.,][]{kwa11} that are currently being dealt with by assumption and approximation.

\begin{figure}[!ht]
\begin{center}
\includegraphics[height=0.4\textheight]{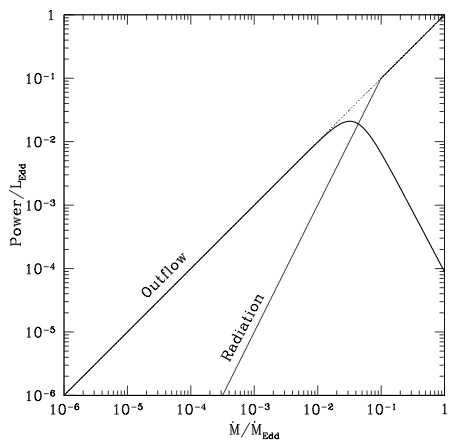}
\caption{Jet power and radiated power in units of the Eddington limit
  \vs{} accretion rate in Eddington units \citep[from][]{css05}.  This
  is a schematic illustration of the transition from low radiative
  efficiency and high mechanical efficiency at low accretion rates to
  high radiative efficiency and low mechanical efficiency at high
  accretion rates.}
\label{fig:jetvsrad}
\end{center}
\end{figure}

Unlike radio AGN, quasars and high nebular excitation AGN release energy primarily
in the form of radiation generated by %only ``viscous'' in the broad
%sense 
dissipation in a thin accretion
disk \citep{ss73}.  Radio AGN in clusters and normal ellipticals emit
the bulk of their energy not in radiation but in a mechanical form
that is accompanied by radio jets and lobes.  This form of energy output is thought to
be generated by radiatively inefficient accretion from a high
pressure, optically thin but geometrically thick, ionized disk (see
\citealt{nm08} for a review).  The ions and electrons are out of
thermal equilibrium, and their cooling times exceed the flow time onto
the SMBH.  Therefore, most of the gravitational energy released by the
inflow cannot be radiated.  The energy is instead carried inward with
the accretion flow, although some may be driven outward in a
mechanically-dominated jet and/or a wind, i.e., ADIOS  \citep{bb99}.  Because so
little radiation emerges from the accretion flow, they have been dubbed radiatively inefficient
accretion flows (RIAF) or advection dominated accretion flows (ADAF).

With the known exceptions of a few distant cooling
clusters hosting central quasars \citep{rfs10, clf99, sba10},  
BCGs and ellipticals host RIAFS.  Their nuclear X-ray and
radio synchrotron luminosities lie below a few percent of their jet
powers \citep{sce00, mh07, hf11, fr95}.  By analogy with black hole
binaries \citep{fct99, mhd05, krk08}, a quasar or radiation dominated
accretion flow forms when the physical accretion rate exceeds a few
percent of the Eddington accretion rate \citep{bbr84, fkm04, mhd05,
  css05}.  The Eddington rate can be expressed as $\dot M_{\rm Edd} =
2.2\epsilon^{-1}M_{\rm BH,9}\rm\ \msun\ yr^{-1}$, where $\epsilon$ is
the conversion efficiency between accretion power and radiated power,
which is generally assumed to be $\sim 10$ percent, and $M_{\rm BH,9}$
is the black hole mass in units of $10^9M_\odot$.  For accretion far
below the Eddington rate, a high pressure RIAF forms \citep{ny94,
  bb99}.  As shown here schematically in Fig.~\ref{fig:jetvsrad} 
\citep{css05}, AGN power is dominated by mechanical outflows when $\dot
M/\dot M_{\rm Edd} \ll 1$.  When $\dot M/\dot M_{\rm Edd}$ exceeds a
few percent, the AGN transitions to a quasar and it releases the
binding energy of accretion in the form of radiation.
 
Although the black hole masses in  BCGs are all but unknown, 
their status as the largest galaxies in the Universe
ensures that they harbor massive black holes \citep{mmg11}. 
%the evidence in support of this picture is compelling.
For jet powers  $P_{\rm jet} = 0.1 \dot Mc^2$, accretion rates implied by
mechanical jet powers lie in the range $10^{-4}$ -- $10^{-2}$ of the
Eddington rate \citep{rmn06, hjf09, hf11}, which is well within the RIAF regime.
Physical accretion rates typically correspond to several hundredths of
a solar mass per year.  Only in extreme cases, such as MS0735 and
Hydra A, do the implied accretion rates approach or exceed $1 ~M_\odot
~\rm yr^{-1}$.  The combination of relatively high jet powers and
radiatively inefficient accretion is consistent with nuclear black
hole masses exceeding $\sim 10^9~M_\odot$.  In this picture,
quasars at the centers of clusters \citep[\eg,][]{clf99} are important because they
are accreting at high Eddington rates 
and may be in transition between radiatively efficient and inefficient AGN (i.e., 
quasar mode and radio mode). 
%\citep{hf11}.

\subsection{Hot Bondi Accretion or Cold Molecular Accretion?} \label{sec:bondi}
 
Fueling AGN by spherical Bondi accretion from hot atmospheres is appealing for
several reasons.  The Bondi mechanism is simple.  The accretion rate is
related to the temperature and density of the atmosphere
at the location of the Bondi radius which is five or six decades larger than the 
event horizon.   Moreover, Bondi accretion provides a simple feedback
mechanism because the Bondi accretion rate for $\gamma = 5/3$ depends
on the gas properties only through the entropy index, $\entind = kT /
\nelec^{2/3}$,  which is the gas property most directly affected by heating and
cooling.   
%The Bondi radius has been nearly resolved in some galaxies by \chandra{}
%allowing these parameters to be constrained.
% Bondi accretion can be treated analytically.  It
% operates on scale lengths six to seven decades larger than the scale
% of the event horizon, so it is easier to model.  ****NOT REALLY***
Most importantly, hot atmospheres are naturally able to provide a
steady supply of fuel when the gas is dense
enough (\ie, the entropy is low enough) to provide an accretion
rate that is sufficient to power the AGN.  The latter condition is probably
met for low power jets \citep{adf06}.  However, host galaxies of jets
exceeding $\sim 10^{45} ~\rm erg~s^{-1}$ would have great difficulty
accreting enough hot fuel to power them \citep{rmn06, hec07, mrn11}.
There are other difficulties.  Bondi accretion is expected to operate
without producing a substantial reservoir of cold gas.  While this
would be consistent with giant elliptical galaxies, it is at variance
with the levels of cold gas found in BCGs \citep{e01,
  sc03, eom10, ddo11, obg10}.  If Bondi accretion is operating in low
power AGN and cold accretion in high power AGN, the two feedback
mechanisms must operate seamlessly on hot atmospheres spanning seven
decades in jet power and X-ray luminosity.  With rare exceptions, enough cold molecular gas is
available in BCGs and probably in normal ellipticals to power their
AGN \citep{e01, soker08}.  Sufficiency is no proof, but feeding the
AGN with cold gas would obviate the need for two modes of accretion.

Further complicating matters, it remains unclear how the accretion
rate for hot gas is affected by the small amount of angular momentum
likely to be present \citep{nf11, ps10}.  Using 2-dimensional
magnetohydrodynamic simulations, \citet{pb03} concluded that the
accretion rate for slowly rotating hot gas could be orders of
magnitude less than the Bondi accretion rate.  In their model, angular
momentum is transported outward by magnetorotational instability
\citep[MRI,][]{bh98}.  More recently, \citet{is10} have made a
hydrodynamic model for the accretion flow that includes thermal
conduction and bremsstrahlung radiation losses.  With no angular
momentum transport, the ``centrifugal barrier'' plays a key role in
this model, and it is no surprise that they also find accretion rates
much smaller than the Bondi rate.  By contrast, assuming a kinematic
shear viscosity of the form $\alpha c_{\rm s} r$, where $c_{\rm s}$ is
the isothermal sound speed and $r$ is the radius, \citet{nf11} found
that slowly rotating gas can accrete at close to the Bondi rate if
$\alpha \sim 0.1$.  The incompatibility of these results reflects
widely differing physical assumptions concerning the transport of
angular momentum.  Their validity depends critically on the highly
uncertain transport properties of the plasma, particularly the
effective viscosity (section \ref{sec:transport}).

New evidence for a rising temperature gradient into the nucleus of NGC
3115 is consistent with Bondi flow \citep{wiy11}, but it is also
consistent with models having much lower accretion rates for the hot
gas.  For the Bondi accretion rate, $\mdotbondi$, \citet{wiy11}
estimate that the X-ray luminosity of the AGN in NGC~3115 is at least
six orders of magnitude smaller than $0.1 \mdotbondi c^2$.  That also
makes the nuclear X-ray luminosity at least an order of magnitude
smaller than the thermal power, $\mdotbondi 5 k T/ (2 \mu m_{\rm H})$,
conveyed inward through the Bondi radius by the gas.  If Bondi
accretion does work, the radiative inefficiency of NGC~3115 is truly
remarkable.

\subsection{Black Hole Spin and AGN Feedback} \label{sec:spin}

A black hole's spin encodes its formation history \citep{wc95, ms96,
  ssl07, vs11}.  A black hole growing by equal mass black hole mergers
or during extended periods of prograde gas accretion is expected to
be endowed with a high spin parameter \citep{vmq05}.  Black holes
growing through the accretion of smaller black holes and by accretion
of randomly-oriented gas disks are expected to have low spin
parameters \citep{kp06, kph08, hb03, gsm04}.  

Like all astrophysical objects, black holes
are almost certainly spinning.  But estimates of their spin parameters are generally indirect
and model dependent.  A connection between
a black hole's rotation rate and its energetic output has been inferred
from an apparent dichotomy between
rapidly-spinning, radio loud AGN and slowly spinning, 
radio quiet AGN \citep{ssl07, msr11,wcw11}.  However, the association
between instantaneous jet power and rotation rate depends on several factors 
including, the relationship between total radio power and its core radio luminosity, and the assumed black hole mass
and its accretion disk geometry \citep{tnm10, bf11}.  In other words, only an indirect connection
can be made between the angular momentum of a black hole and its emergent properties. \footnote[1]{A correlation between radio power and black hole spin was recently found for
stellar mass black holes by \citet{nm12}.}
More direct estimates of  black hole spin parameters have been obtained for several
AGN using the shape and
amplitude of the relativistically-broadened, 6.4 keV fluorescent  iron
feature.  These measurements have generally revealed high spin parameters \citep{miller07, tnf95, fnr95, brn11}.
%are consistent with high spin parameters in black holes exceeding
%$10^8~M_\odot$.  This method relies on difficult measurements of the
%line shape over a broad region of the X-ray spectrum, and is subject
%to significant systematic uncertainties from foreground absorption.
%Nevertheless, recent results provide promising indications that some
%SMBHs spin rapidly [ED].
In an exciting new development, long baseline millimeter interferometric
observations are beginning to image the event horizon silhouettes in M87 and
Sgr A* \citep{dwr08, fdb11,  blr11}.  The spin parameter is derived from the 
structure of the silhouette.  The initial results indicate rapidly and slowly spinning black holes for M87 and Sgr A*, respectively. Due to resolution limitations, this method will
be accessible only to M87 and Sgr A* for the foreseeable future.

Spin may play an important role in AGN feedback.  The rotational
energy of a maximally-spinning, $10^9 ~M_\odot$ black hole exceeds
$10^{62}$ erg.  This spin energy is significant with respect to the thermal
energy of its surrounding X-ray atmosphere and is enough to quench a
cooling flow for several Gyr.  Under the right circumstances, the
rotational energy may be released in the form of jets through the
Penrose-Blandford-Znajek mechanism \citep{bz77, hk06}.
%% Note: process -> mechanism.  Penrose recognised that spin energy
%% could be tapped, but his ``process'' for extracting energy does not
%% apply here.
The jetted outflow is coupled to the black hole's rotational energy by
magnetic fields supplied by and anchored to an accretion disk
\citep{meier99, bhk09,kh10,khh05}.  Tapping into spin to power a
jet requires a significant level of accretion.  The coupling between
the angular momentum of the black hole and the accretion disk is weak for low
spin parameters and may depend on whether the accretion disk is in
pro- or retrograde rotation \citep{ges10}.  It is unclear whether jet
power is derived primarily from the spin itself or from the binding
energy released by accretion.  But in the context of feedback, it is
accretion that must couple the rotational energy of the black hole to
the feedback loop itself \citep{mrn11}.

We became interested in the possibility of coupling spin power to a
feedback loop after noticing that some systems with extraordinarily powerful jets,
such as those in
the MS0735 cluster, are hosted by BCGs with relatively little cold gas
\citep{mkr09, mrn11}.  If powered by accretion alone, MS0735's $\sim
10^{62}$ erg AGN (Fig.~\ref{fig:ms07fe}) must have accreted at a rate
of several solar masses per year for the $\sim 10^8$ yr duration of
its outburst.  This implies that its SMBH accreted several
$10^8~M_\odot$ of gas.  Yet its BCG harbors less than a few
$10^9~M_\odot$ of cold gas \citep{sc08} and shows no sign of star formation.  If powered by accretion, an uncomfortably
large fraction of its gas shed its angular momentum and was dispatched
onto the black hole without forming stars.  This
scenario seems unrealistic.  If the jets formed instead through a modified BZ
process \citep{nbb07}, even a rapidly-spinning, $10^9~\rm
M_\odot$ black hole would require a high 
% Likely hosting a $10^9~M_\odot$ or larger
% black hole, the BZ process \citep{nbb07} would require a high
accretion rate to power its jet, which is the scenario we hoped to avoid by appealing to spin.  Note that accretion
alone is unable to generate power above a few tens of percent of $\dot
Mc^2$.  Therefore, an AGN substantially in excess of this limit must be powered by spin or 
by a process other than accretion.  

A mechanism able to access the rotational energy of a SMBH efficiently and at low accretion rates, \ie, $P_{\rm jet} > \dot Mc^2$, may be
needed to sustain the most powerful AGN \citep{gtf10, punsly11, fjr11, mkr09, mrn11}.
One such mechanism may be the magnetically arrested accretion model of \citet{tnk11}.
Their model produced
jet efficiencies exceeding 100\%, i.e., $P_{\rm jet} > \dot M c^2 $, from a rapidly spinning
SMBH at the center of a disk threaded by a large-scale poloidal magnetic field.
High magnetic field pressure is sustained in part by ram pressure from accreting gas (see also \citealt{cao11, mtb12}).   Accreting matter presses the poloidal disk field into the ergosphere,
maximizing energy yield via the Blandford-Znajek mechanism.
%Efficiencies exceeding 100\% are
%impossible unless energy is tapped from the spin of the black hole. 
%The key to magnetically arrested accretion is a large scale poloidal magnetic field.  
It has been reasoned
that shear in the disk causes the toroidal component of the field to
dominate anywhere close to the black hole.  In particular, if
accretion is governed by turbulent viscosity generated by MRI
\citep{bh98, dhk03}, a large-scale poloidal field seems unlikely.
However, poor understanding of the accretion process leaves plenty of
room for a mechanism to obtain strong poloidal fields.  Such a
mechanism is more plausible for hot, pressure supported,
quasi-spherical accretion flows than for cold accretion through a thin
disk.
%consistent with higher jet efficiencies at low accretion rates.

Spinning black holes launch Poynting flux jets along the
black hole's spin axis \citep{ntl08} that carry
very little matter with them  \citep{hk06}. 
Using the distribution of X-ray cavity sizes as a function of their distance
from the nucleus, \citet{dlf08} argued that their size distribution is
consistent with predictions for Poynting jets but not for adiabatically
expanding hydrodynamic cavities.  On the other hand, the large jet
powers inferred from X-ray cavities require energetic protons or other
heavy particles to maintain pressure balance.
If jets are dominated by Poynting flux on small scales, they must undergo significant mass loading
before reaching the radio lobes.

\section{AGN Heating in Distant Clusters}

Most of what we have learned about AGN heating is
based on X-ray observations of bright, nearby clusters that are much easier to observe.  
Atmospheric heating by AGN is expected to become increasingly important in distant
clusters as AGN activity rises.   
The excess entropy in hot atmospheres \citep{v05, pcb11} may be the
relic of primeval heating by AGN and star formation \citep{k91}.  But whether the excess entropy was injected in an early ``preheating''
phase \citep{k91} or more gradually over time is unknown. 
Systematic studies of atmospheric heating in distant clusters
are needed to address this issue.

\begin{figure}[!ht]
\begin{center}
\includegraphics[height=0.25\textheight]{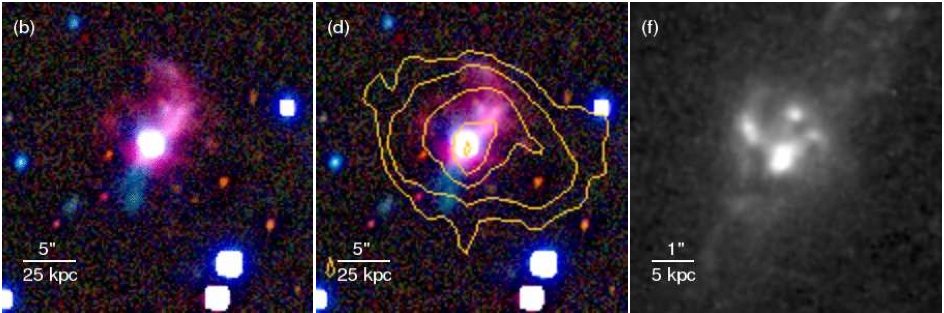}
\caption{Optical structure of the BCG of MACS J1931.8-2634. 
  SuprimeCam BRz image of the central 30 arcsec by 30 arcsec. (b): 
  the contribution from the old stellar population of the
  BCG (as traced by the SuprimeCam I-band image) was subtracted from
  each of the B,R and z images before combining them to create a color
  image. This enhances the blue and pink features visible to the
  southeast and northwest of the central AGN. ÒPinkÓ signals
  contributions from predominantly the blue (B) and the red (z)
  channel. At the redshift of the cluster, the H$\alpha$ line falls into the
  z-band, and thus this emission likely stems from H$\alpha$ nebulosity
  surrounding MACS J1931.8-2634. The blue emission
  signals a young stellar population (d):
  Overlay of the radio emission on (b). (f): The central 7.5 arcsec 
by  7.5 arcsec of an HST snapshot. Note the bright knots in a
  spiral-like structure emanating to the Northwest of the
  brightest central knot \citep{eav11}. 
}
\label{fig:macsbcg}
\end{center}
\end{figure}

Examples of the few targeted studies of distant cavity systems are the
$z=0.35$ clusters RBS 797 \citep[and Doria et al. in
  preparation]{scd01, cmw11} and MACSJ1931.8-2634 \citep{eav11}.  Both
have AGN mechanical powers in the range $10^{45}$ -- $10^{46}~\rm erg
s^{-1}$.  The MACSJ1931 BCG \citep{eav11} is a spectacular galaxy with
knots of star formation, nebular emission, radio jets, and X-ray
cavities (Fig.~\ref{fig:macsbcg}).  After accounting for AGN heating,
its star formation and cooling rates at $170 ~\rm M_\odot ~yr^{-1}$
are both consistent with each other. This is a solid example of a
burgeoning BCG that may be regulated by AGN feedback. \citet{hfe11}
examined the prevalence of X-ray cavities in a sample of 83 MACS
clusters lying in the redshift range $z=0.3-0.7$.  They found no evolution in the
average jet power or cavity properties over their sample.  This study shows that radio-AGN feedback was operating on
cooling flows in the most massive clusters when the Universe was nearly  half its current age as it is
operating today.  
%Only nuclear X-ray emission increases with  redshift \citep{hfe11}.
%However, such luminous, cooling flow clusters are not necessarily
%representitive of normal BCGs at high redshift.

Other studies of AGN heating beyond a redshift of $0.3$ have
approached this problem statistically. Most have focussed on the
detection frequency of AGN in the general population of galaxies in
clusters \citep{mkm02, msm09, ems07, gse09}.  These studies found a
dramatic increase in the numbers of X-ray and infrared-emitting AGN in
galaxies projected within clusters out to and beyond
redshift one.  This increase reflects the universal rise of AGN
activity with increasing redshift, but it does not probe atmospheric
heating because radiation couples weakly to the keV gas.

Radio AGN are strongly coupled to the keV gas and are thus able to probe atmospheric heating.
Radio AGN in clusters lying within $z\sim 1$ tend to be centrally
concentrated and are more numerous than X-ray and infrared AGN
\citep{bvk07, gse09}.  \citet{gse09} and \citet{mmn11} found that the detection
fraction of radio AGN rises with redshift by less than a factor of two
in the range $z\sim 0.2-1$, while the numbers of X-ray AGN rise
dramatically over this redshift range \citep{gse09}.  \citet{hsh09, hse11} measured the incidence of radio AGN in a
dozen X-ray clusters in the range $z\sim 0.4-1.2$, specifically
chosen to match Coma-like progenitors.
They found roughly 17 cluster radio galaxies in their sample within a
projected radius of 1 Mpc from the cluster centers.  Unlike
\citet{gse09} and \citet{mmn11},  \citet{hse11}
found strong redshift evolution of the cluster radio luminosity
function, consistent with a near tenfold increase in the AGN heating
rate by galaxies throughout clusters at a redshift of one compared to
today.  

\citet{mmn11} cross correlated the NVSS radio catalog with 242
clusters from the 400 Square Degree X-ray cluster catalog
\citep{bvh07}.  Their
clusters lie within the redshift  range $0.2<z<0.7$ and have  bolometric
X-ray luminosities of $10^{43} <L_{\rm x} <10^{45}$ erg s$^{-1}$.
Ma found a high incidence of radio galaxies
within a projected radius of 250 kpc indicating that AGN heating is probably
significant.  
%However, lacking deep X-ray
%imagery, surveys of radio AGN heating, e.g., \citep{bvk07} must rely on scaling
%relations between radio luminosity and mechanical power \citep{brm04,
%  bmn08,cmn10} to estimate AGN heating rates.   
Using scaling relations between radio luminosity and cavity power
\citep{bmn08, cmn10}, Ma \etal~found that the integrated AGN power
over this luminosity range is nearly constant and independent of
cluster X-ray luminosity.  Only a mild, perhaps two-fold increase in
average jet power is seen between $0.15 <z<0.6$, which is consistent
with  the results of \citet{gse09}. The key result is that the ratio of jet power to
X-ray luminosity increases fourfold, from roughly $1/2$ in the most luminous,
$L_{\rm x} \sim 10^{45}~\rm erg ~ s^{-1}$, clusters  to $\simeq 2$ in lower
luminosity, $L_{\rm x}
\sim 5 \times 10^{43}~\rm erg ~ s^{-1}$, clusters.  The relatively constant AGN power
input with X-ray luminosity implies that AGN heating is more effective in
lower luminosity clusters containing fewer gas particles than high luminosity clusters.  
AGN heating in groups may be more dramatic,  in some cases
sweeping away their hot atmospheres \citep{gsf10}.   Finally, radio AGN heating of the hot atmospheres of elliptical galaxies was apparently operating at $z\sim 1.2$ 
as it is today \citep{dla12}.

Assuming no evolution, \citet{mmn11} found an average energy input of
approximately $0.2$ keV per baryon within $R_{500}$.  This value would
double if the integrated jet power remained constant out to $z=2$, and
would increase even more if the radio luminosity function increases in
clusters at higher redshifts.  More recent work by Ma and collaborators
studying a much larger sample of X-ray clusters indicates that AGN
heating approaches $\sim 1$ keV per baryon in low luminosity clusters.
AGN heating would then be significant
compared to the $\sim 1$ keV per baryon needed to explain the excess
entropy in clusters \citep{k91, wfn00, v05}.  These studies taken
together imply that gradual heating over time may be as important or
more important than Kaiser's early preheating scenario
\citep[\cf][]{yts11}. 

%Distant clusters QSOs \citep{crawfQSO}
\subsection{Are Strong Cooling Flows in Ascendency?}

A potential consequence of AGN heating in distant clusters would be
the growing, albeit tentative, evidence for a decline in the frequency
of {\it strong} cooling cores (flows) in distant clusters.  Despite an
expected tendency for X-ray-selected clusters to be biased in favor of
bright, cooling cores \citep{emp11}, distant clusters in the 400SD,
WARPS, and RDCS surveys lack the large numbers of bright central cusps
of X-ray emission associated with cooling flows 
in nearby samples \citep{vbf07, str10},  such as the \textit{ROSAT} All Sky Survey
\citep{cae99}.
%Using high resolution \chandra{} X-ray observations of clusters from
%the 400 Square Degree, RDCS, and WARPS surveys, \citet{str10} also
%found a shortage of high surface brightness cusps associated with
%strong cooling flows at a median redshift of 0.83, although the
%evolution they find is less dramatic than suggested by \citet{vbf07}.
Bolstering this finding, the incidence of radio loud BCGs in 400SD
clusters is roughly 30\% \citep{mmn11}, a rate that is consistent with
the detection fraction of relatively normal Sloan clusters
\citep{bvk07}, but far below the $70\%-90\%$ detection rate in strong
cooling flow clusters \citep{b90, bvk07, cdv08, mhr09, df08}.
% In the X-ray flux-limited brightest 55 sample from the Rosat All Sky
%Survey, \citet{df08} found that the radio AGN in essentially all
%strong cooling flows were ``on,'' showing X-ray cavities and radio
%emission.

The radio detection rate found by \citet{mmn11} is consistent with the
absence of {\it strong} cooling flows in distant 400 Square Degree
X-ray clusters, but not inconsistent with significant numbers of smaller
cooling flows.  A similar conclusion was reached by \citet{smv11}, who
found that strong nebular line emission features were absent in 77 BCGs selected
from the 160SD cluster survey.  Nebular emission is a
reliable indicator of the presence of a cooling flow \citep{chj83,
  hbv89, dsg92, cae99, mvr10}, yet \citet{smv11} found their numbers declining at
 redshifts $0.3-0.5$.  The \citet{smv11} result was recently confirmed by \citet{m11} using
  a different sample of clusters.  
  
These studies do not preclude
the existence of moderate to weak cooling flows in distant clusters.
But they are consistent with declining numbers
of {\it strong}, Perseus-like cooling flows at higher redshift.  Andy Fabian
has suggested to us that this decline may reflect the bias against the X-ray detection of
cooling flow clusters hosting central X-ray-bright quasars that outshine the 
thermal emission from the surrounding cluster \citep[e.g.,][]{rfs10, sba10}.  This must
be happening at some level.  But the numbers of X-ray-bright AGN in
distant BCGs are unlikely to be high enough to account for this effect.  Although selection
effects are always an issue  \citep[e.g.,][]{str10}, the evidence for declining
numbers of strong cooling flows beyond redshift $\sim 0.3 $ is growing and is probably real.

\section{Concluding Remarks}  \label{sec:conclusions}

A strong empirical case has emerged in recent years that AGN feedback limits gas cooling and star
formation in the cores of low redshift galaxy clusters and groups.  The impact of AGN outbursts on cluster atmospheres is evident
in high resolution X-ray images.  They displace gas, drive shocks and
sound waves, and they transport low entropy gas and heavy elements
outward from cluster cores.  AGN in BCGs also respond to their
environments with a higher incidence of radio activity and greater jet
powers in richer clusters.  The average mechanical power output
adjusts to the level required to inhibit gas from cooling and forming
stars.  While cooling hot gas may not be the only cause of AGN
activity in BCGs, it is clearly implicated as a significant factor.
Star formation and cool gas are found exclusively in BCGs located in cluster cores with the
shortest cooling times that are most prone to thermal instability.

Understanding of the AGN mechanical feedback loop lags the
observational evidence for it.  A number of processes have been
identified that can play some role in dissipating and distributing
outburst energy in hosting atmospheres, but it remains unclear which
of them are most significant.  Perhaps the greatest obstacle to
progress here is poor understanding of transport processes.
Certainly, anisotropic transport due to magnetic fields must be taken
into account, but it is possible that transport and dissipation in the
intracluster plasma are no less complex than they are in the solar
wind plasma, in which case existing models are inadequate.  Further
work is required to clarify this issue.

Bondi accretion could power most AGN outbursts observed locally, but
small angular momentum may cause a drastic reduction in accretion
rates, another issue remaining to be resolved.  There is a small
number of systems in which the fuel supply from Bondi accretion is
insufficient to power the jets.  Large quantities of cold gas found in
many systems might provide an alternative source of fuel in those
cases.  It remains to be determined which source of fuel, hot gas or
cooled hot gas, is mainly responsible for powering mechanical feedback
in BCGs.  Furthermore, the enormous energies of some outbursts
challenge all accretion powered models, suggesting that black hole
spin may need to be tapped as the energy source in at least some
cases.  General relativistic magnetohydrodynamical simulations are just beginning to shed light on this issue.

Observations with \chandra{} and \xmm{} have been critical to progress
in this field so far.  In the near future, it will be possible to
address many of the open questions with a suite of new instruments.
The planned X-ray survey of \textit{eROSITA} will provide a census of
systems with extended hot atmospheres, including large samples of
galaxy clusters at redshifts beyond one.  Many of these and other
nearby bright clusters will be observed with the microcalorimeter array
on the ASTRO-H observatory, hopefully revealing 
the level of turbulence in the hot gas produced by AGN outflows and
other processes.  The \textit{eROSITA} survey will be
complemented by forthcoming radio surveys.  Notably, the planned ASKAP
EMU and Westerbork Wodan surveys will cover the full radio sky at
$\sim 1$ GHz, with much greater sensitivity and spatial resolution
than the existing NVSS.  At lower radio frequencies, LOFAR will be an
excellent tool for studying older radio outbursts that have faded at
higher radio frequencies.  Surveys by these instruments will be
particularly valuable for studying mechanical outbursts from AGN at
redshifts of one and beyond to investigate the history of mechanical
feedback during the formation of groups and clusters.  They will
also provide data on mechanical feedback in the era when the radiation
powered quasar mode of AGN feedback is thought to have played a much
greater role than it does today.  The high sensitivity and angular
resolution of ALMA will enable detailed studies of the distribution
and dynamics of cold gas in elliptical galaxies, and the relative importance of
star formation and starburst winds in regulating the flow of gas onto the nucleus. 
These exciting new measurements will help to
constrain the origin, perhaps as cooled hot gas, and fate of the
molecular gas, in star formation and fueling AGN.  Measurements of gas
dynamics with ALMA will also provide much better constraints on black
hole masses in elliptical galaxies, hence their Bondi and Eddington
accretion rates.

\acknowledgments

We thank Helen Russell, Massimo Gaspari, Christoph Pfrommer,
Prateek Sharma, Bill Mathews and especially Mark Voit and Peter Mendygral
for their help and advice.
This work was supported in part by NASA contract NAS8-03060,  Chandra Large Project Grant GO9-0140X, and generous funding from the Natural Sciences and Engineering Research Council of Canada, and an SSEP grant from the Canadian Space Agency.

\bibliographystyle{apj}

\bibliography{review_revised}{}

\end{document}